\documentclass[%
 reprint,
 amsmath,amssymb,
 aps,
]{revtex4}

\usepackage{graphicx}
\usepackage{dcolumn}
\usepackage{bm}

\usepackage{hhline}
\usepackage{setspace}

\usepackage{hyperref}
\usepackage{placeins}
\usepackage{subcaption}
\usepackage{orcidlink}

\usepackage{tikz}
\usetikzlibrary{arrows.meta, positioning}

\begin{document}

\preprint{APS/123-QED}
\def\mean#1{\left< #1 \right>}

\title{First Experimental Demonstration of Reinforcement Learning-Based Tuning on the PSI Injector 2 Cyclotron}

\author{M.~Haj~Tahar \orcidlink{0000-0003-2995-2868}}
\email{malek@transmutex.com}
\affiliation{Transmutex SA, Vernier, Switzerland}

\author{W.~Joho}
\affiliation{Transmutex SA, Vernier, Switzerland}

\author{E.~Solodko}
\affiliation{Transmutex SA, Vernier, Switzerland}

\author{M.~Bocchio}
\affiliation{Transmutex SA, Vernier, Switzerland}

\author{S.~Marquie}
\affiliation{Transmutex SA, Vernier, Switzerland}

\author{M.~Busch}
\affiliation{Transmutex SA, Vernier, Switzerland}

\author{A.~Barchetti}
\affiliation{Paul Scherrer Institut, Villigen, Switzerland}

\author{J.~Grillenberger}
\affiliation{Paul Scherrer Institut, Villigen, Switzerland}

\author{J.~Snuverink \orcidlink{0000-0002-1455-3226}}
\affiliation{Paul Scherrer Institut, Villigen, Switzerland}

\author{M.~Schneider}
\affiliation{Paul Scherrer Institut, Villigen, Switzerland}

\begin{abstract}
Reliable operation of high-power proton cyclotrons is a critical requirement for Accelerator Driven Systems (ADS) and other large-scale applications. Beam tuning in such machines is traditionally performed manually, a process that can be slow, non-optimal, and difficult to execute in the presence of faults or changing conditions. To address this, we developed and deployed a machine learning (ML) based tuning framework on the Injector 2 cyclotron at PSI, chosen as an ideal testbed for high-power operation. The system combined a tailored reinforcement learning (RL) algorithm with real-time diagnostics and control, and incorporated accelerator-physics inspired adaptations such as an overshoot strategy that reduced magnetic field settling times by nearly a factor of six. Over an extensive 12-day operational test campaign, relatively long in the context of real-time ML experiments, the RL agent successfully tuned the machine across multiple operating points. For each investigated configuration, stable policies were obtained within a few hours of online training and subsequently demonstrated reliable low-loss operation during overnight evaluation runs. Crucially, the learned policy remained effective when transferred from low-current training to operation at beam currents up to 800 µA, demonstrating robust generalization under appropriately adapted operational constraints. These results constitute the first demonstration of RL-assisted tuning on a high-power cyclotron, with direct relevance to ADS-class drivers.
\end{abstract}

\pacs{Valid PACS appear here}
\maketitle


\section{Introduction}

The High Intensity Proton Accelerator (HIPA) complex at the Paul Scherrer Institut (PSI) has been in operation since 1974, and today delivers the most intense continuous-wave proton beam worldwide, with an average power of 1.42 MW produced by a two-stage cyclotron system \cite{grillenberger2021high}. In the first stage, Injector 2 cyclotron accelerates protons from 0.87~MeV, provided by a Cockcroft–Walton DC linear accelerator, up to 72 MeV. The second stage, the 590 MeV ring cyclotron, then boosts the beam to its final energy, with a maximum average current of 2.4 mA. After extraction, which is achieved with efficiencies exceeding 99.98\%, the beam is transported to multiple target stations for meson and neutron production \cite{kiselev2021meson}, enabling world-class programs in particle, nuclear, and materials science. 

Cyclotron-based proton drivers, as demonstrated at PSI, combine compactness, high energy efficiency, and excellent beam availability, making them attractive candidates for future Accelerator-Driven Systems (ADS) \cite{rubbia1995conceptual,abderrahim2012myrrha}. ADS couple a high-power proton accelerator to a subcritical nuclear reactor for applications such as nuclear waste transmutation and sustainable energy production. Because the reactor power depends directly on the external proton beam, these facilities impose exceptionally stringent reliability requirements on the accelerator, demanding beam interruptions that are orders of magnitude less frequent and significantly shorter than those tolerated in existing high-power accelerator facilities.

Within this context, Transmutex SA is developing an ADS concept aimed at generating clean energy while simultaneously reducing the long-term radiotoxicity of nuclear waste. The architectural similarities between the HIPA cyclotrons and the envisioned ADS driver make PSI’s Injector 2 cyclotron an ideal testbed to investigate the control strategies and automation capabilities required for such systems.

At PSI, proof-of-principle experiments have already demonstrated that, following the failure of a Radio-Frequency (RF) cavity in the 590 MeV ring cyclotron, beam delivery can be recovered within several minutes by retuning the remaining cavities~\cite{HajTahar_FFA2024}. Although this demonstrates the feasibility of fault compensation, recovery on the order of minutes remains insufficient for ADS operation, where compensation must occur within only a few seconds. More generally, even during routine accelerator operation, slow drifts caused by temperature variations, magnet hysteresis, beam loading, and other environmental effects require continuous retuning to maintain optimal beam quality. These considerations motivate the development of autonomous control strategies capable of compensating both routine operational drifts and, ultimately, fast fault scenarios.

Machine Learning (ML) has emerged as a promising candidate to address this challenge \cite{annurev-AEdelen}. By leveraging data-driven models and adaptive optimization, ML offers the potential to automate cyclotron tuning, compensate for drifts, and adapt to changing operational conditions in real time. While ML approaches have been explored in linear accelerators and light sources, their application to high-power cyclotrons remains largely unexplored \cite{Edelen2016, Scheinker2020, Hirlaender2023}. 

The choice of reinforcement learning in the present work is motivated by the nonlinear and operating-point-dependent nature of cyclotron beam dynamics. As shown later in this paper, the response of the beam phase to variations in main coil current, RF voltage, and trim-coil currents depends strongly on the turn number, resonator configuration, and machine working point. Consequently, a single fixed Jacobian or response-matrix controller is only locally valid, and operation across multiple configurations would require repeated system identification, gain scheduling, or retuning by operators. Reinforcement learning was therefore investigated as a state-dependent control approach capable of directly mapping machine observations to corrective actions while respecting operational constraints and safety limits.

To fill this gap, PSI Injector 2 cyclotron was chosen as an experimental testbed, combining the complexity of a high-power machine with the flexibility needed for R\&D.

The present work reports on the first experimental demonstration of RL-assisted tuning on Injector 2 cyclotron. A tailored Reinforcement Learning (RL) framework was integrated with real-time diagnostics and control of the machine, enabling the agent not only to converge on optimal settings within hours but also to maintain stable operation during overnight evaluation runs by autonomously compensating drifts. This effort was carried out over an extensive 12-day experimental campaign, which is remarkably long in the context of real-time ML control studies. The results of this campaign provide an essential step towards establishing RL-based control as a reliable tool for fault compensation and robust operation in ADS-class drivers.

To move beyond proof-of-concept and towards reliable deployment, the Injector 2 campaign was carefully crafted to address a set of fundamental questions. At the technical level, these include identifying which actuators and diagnostics provide the greatest leverage for beam stabilization, how the machine should be initialized before handing control to an RL agent, and what forms of surrogate modeling or pretraining are most effective in accelerating convergence. At the methodological level, the campaign explored how quickly an RL policy can converge in a real cyclotron, whether a model trained at low current generalizes to higher-intensity operation, how frequently retraining is required to counteract drifts, and to what extent transfer learning across turn numbers can reduce setup time. At the heart of this effort lies the central question: can a cyclotron be operated reliably with limited human intervention, using ML-based control strategies that meet the stringent demands of ADS-class drivers?

The remainder of this paper is organized as follows. Section II describes the Injector~2 cyclotron and its diagnostics, together with the control framework used for the experiments. Section III presents an analysis of historical data and feature engineering choices that informed the RL environment design. Section IV outlines the overall tuning strategy and campaign design. In Section V, the turn- and current-dependent beam dynamics are characterized using radial probe and Jacobian measurements, providing the physics context for the tuning challenge. Section VI discusses the adaptations required to integrate accelerator physics constraints into the ML framework, including the overshoot strategy and reward shaping; in particular, the overshoot strategy was found to accelerate the tuning process by nearly a factor of six. Section VII details the RL methodology employed, while Section VIII reports the experimental results, covering low-current training, overnight evaluation, and generalization to higher beam currents. Section IX presents the conclusions and outlines the roadmap toward scaling the demonstrated methodology to the HIPA complex and future ADS-class accelerator drivers.

\section{Experimental Setup}

\begin{figure*}
\centering 
\includegraphics*[width=16cm]{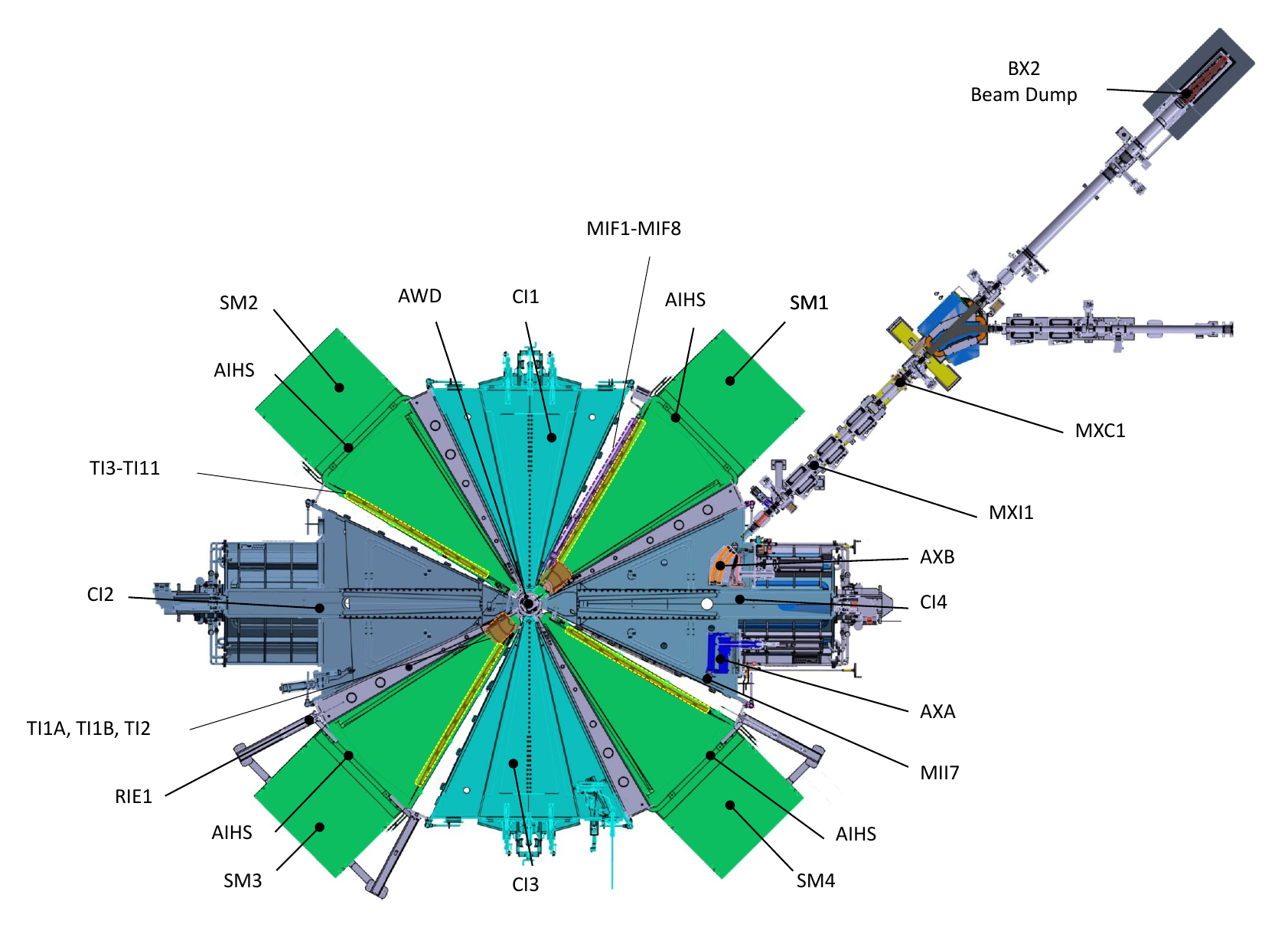}
\caption{Injector 2 Cyclotron experiment setup with key diagnostics.}
\label{fig:inj2_layout}
\end{figure*}

Injector~2 is a four-sector separated-sector cyclotron designed to accelerate protons from an injection energy of 870 keV to a final energy of 72 MeV \cite{stetson1992commissioning, schneider2019upgrade}. Its principal components include four large sector magnets (SM1--SM4), two double-gap RF resonators (CI1 and CI3), and two single-gap resonators (CI2 and CI4). A schematic layout of the machine with the main diagnostics is shown in Fig.~\ref{fig:inj2_layout}. For brevity, the Injector~2 cyclotron is hereafter referred to simply as Injector~2.

For convenience, the principal Injector~2 actuators, diagnostics, sensors, and the corresponding PSI abbreviations used throughout this work are summarized in Table~\ref{tab:abbreviations} in Appendix~\ref{app:heatmap}.

The proton beam enters Injector~2 from the vertical pre-injection line and is bent into the median plane by the AWD dipole magnet. It then passes through a conical injection channel in the first sector magnet (SM1), which is equipped with a dedicated coil to guide the first turn. During the first six turns, a system of collimators minimizes beam halo and ensures a well-centered trajectory. Additional fine control at this stage is provided by trim coils TI1A/B and TI2 located in SM1 and SM3.

\subsection{Sector Magnets, Trim Coils, and Resonators}

Each sector magnet is powered by a pair of main coils (AIHS) connected in series. These coils generate the primary magnetic field and play a central role in shaping the beam trajectory, thereby defining the overall particle orbit. Nine pairs of trim coils (TI3--TI11), mounted above and below the median plane on one side of the pole, provide localized magnetic field corrections that are indispensable for precise orbit control and efficient extraction.

Beam acceleration is achieved by four RF resonators: two double-gap (CI1 and CI3) and two single-gap (CI2 and CI4). CI1 is critical for defining the first-turn trajectory and is therefore kept fixed during tuning. In combination, the resonators determine the per-turn energy gain and the radial position of successive turns. The AIHS main coils, trim coils, and resonator voltages were all made available as controllable channels for the ML agent.

\subsection{Beam Diagnostics}

The following diagnostics were employed during the ML campaign:

\begin{itemize}
    \item MIF probes (MIF1--MIF8): Eight phase probes in the first sector, providing phase information at different beam energies from injection to extraction. These were the primary reward signals.
    \item Loss monitors: Ionization chambers MII7 (before the extraction septum) and MXI1 (downstream), plus the KXAI collimators at the septum entrance.
    \item MXC1: high-resolution current transformer downstream of the septum, monitoring extracted beam intensity.
    \item Radial probe (RIE1): movable probe system used to obtain radial beam profiles in the extraction region (last 7 turns). This is a slow measurement applied for beam characterization rather than in the experimental ML feedback loop.
\end{itemize}

\section{Historical Data Processing and surrogate model}

\subsection{Historical Data and Feature Selection}
Historical operational data recorded by the PSI archival system were used to construct a data-driven surrogate model for offline RL pretraining. Since archive channels operate at different sampling rates, all signals were resampled to a uniform 200 ms interval. Only operating periods corresponding to machine configurations representative of the Injector~2 experimental campaign were retained. Further details of the data extraction and preprocessing are provided in Appendix~\ref{app:data}.

Based on accelerator expertise and preliminary feature-selection studies, a reduced set of variables was retained for subsequent modeling. The selected features included the trim-coil currents (TIs), resonator voltages (CIVs), main correction coil current (AIHS), beam phases (MIF1--MIF8), beam losses (MII7 and MXI1), extracted beam current (MXC1), and aggregated magnet and air temperatures. Additional engineered features, including estimates of the turn number and beam-stability indicators derived from the MXC1 ripple, were incorporated to improve model performance while preserving physical interpretability.

The final dataset contained approximately 27 million samples, corresponding to more than 63 days of Injector~2 operation.

\subsection{Correlation analysis}
Correlation analysis was performed to validate the selected feature set and identify redundant variables. Strong correlations were observed among magnet temperature sensors and among air temperature sensors, justifying their aggregation into representative mean values. The analysis also confirmed the expected dependence of beam phases on resonator voltages and the secondary influence of trim-coil currents on the radial beam position. correlation heatmap is provided in Appendix~\ref{app:heatmap}

\subsection{Surrogate model}
To support offline pretraining of the reinforcement-learning policy, a data-driven surrogate model was developed to approximate the response of the PSI Injector 2 cyclotron from historical machine data. The final surrogate input vector contained 21 variables: 14 actionable variables, consisting of 12 trim-coil currents, the CI3 resonator voltage, and the main-coil current, together with seven observable variables including temperature information and the voltages of other resonators. The output vector contained the eight phase-probe signals MIF1–MIF8. The surrogate therefore learned a mapping from a 21-dimensional machine-state and actuator vector to an 8-dimensional phase-response vector.

All input and output variables were scaled using limits derived from historical data. These limits were obtained from the combined filtered and raw data ranges and then adjusted to physically meaningful model ranges. A held-out dataset was extracted from the middle of each filtered dataset using a central slice of $10^6$ samples per dataset, giving a total of 3.0 million samples. This held-out dataset was further split into validation and test subsets, with one third used for validation and two thirds used for testing. The remaining data were used for training. To balance the larger datasets, the 2023 training data were down-sampled by taking every seventh row, while the 2024-with-cavity training data were down-sampled by taking every fifth row. After this balancing step, the final training set contained approximately 27 million samples.

Several surrogate-model candidates were evaluated, and the final selected model was a Mixture Density Network (MDN). The detailed network architecture, training procedure, feature weighting, and hyperparameters are provided in Appendix~\ref{app:surrogate}. On the held-out test set, the selected surrogate achieved an average test mean-squared error of $2.4\times10^{-5}$ and an average coefficient of determination of $R^2=0.9898$, demonstrating excellent agreement with historical machine data. The surrogate was therefore selected as the offline environment for RL pretraining.

The purpose of the surrogate was not to replace validation on the real accelerator. Instead, it provided a fast, data-driven environment in which the RL policy could be exposed to representative Injector 2 state-action-response relationships before online deployment. This reduced the amount of exploratory interaction required on the machine and allowed different training strategies to be tested before live operation.

\section{Tuning Strategy and Campaign Overview}
The deployment of RL control in Injector~2 was carefully planned to balance safety, diagnostic coverage, and systematic exploration of machine behavior. A beam current of 20~µA was selected as the reference for training, providing reliable phase measurements at sufficiently low beam power to minimize activation risks in case of instability. At this current, the beam size remained small enough to allow wide action exploration, yet the signal quality was adequate to meaningfully characterize losses and phase shifts.

The tuning strategy was organized around two classes of actuators. The main coil power supply current for the magnetic field (AIHS) and the voltage of resonator 3 (CI3V) were identified as the dominant parameters controlling global beam conditions, specifically the radial trajectory and the extraction energy. Both were included in the RL action space, but their operational boundaries were carefully determined in advance by manual scans to ensure sufficient headroom before interlocks. All twelve remaining trim coils were also available to the RL agent, primarily to compensate for local phase deviations. Reward shaping was employed to discourage excessive trim coil usage, ensuring that corrections reflected genuine improvements in global conditions rather than overcompensation.

The campaign targeted five distinct machine configurations, each corresponding to a different total number of revolutions required for the beam to reach the extraction energy (72, 73, 74, 89, and 60 turns). These operating points were obtained by varying the peak voltages of resonators 2 and 4, thereby changing the energy gain per turn and consequently the total number of turns to extraction. The corresponding resonator settings are summarized in Table~\ref{tab:resonator_setup}. These configurations were deliberately selected to span the three operational scenarios of Injector~2: nominal operation with four resonators, reduced operation with three resonators, and a degraded mode with only two resonators. This allowed us to systematically investigate how the RL agent adapts its tuning strategy under both nominal and failure-like conditions.

\begin{table}[htb]
  \caption{Resonator configuration per experimental stage. Listed are the peak voltages of the four resonators (in kVp) for each turn number.}
  \label{tab:resonator_setup}
  \centering
  \begin{tabular}{c c cccc}
    \hline
    Stage & Turn number & \multicolumn{4}{c}{Resonator setup (kVp)} \\
          &             & Res~1 & Res~2 & Res~3 & Res~4 \\
    \hline
    0 & 72 & 430 & 429 & 451 & 0 \\
    1 & 73 & 430 & 401 & 449 & 0 \\
    2 & 74 & 430 & 371 & 448 & 0 \\
    3 & 89 & 430 & 0   & 449 & 0 \\
    4 & 60 & 430 & 428 & 448 & 428 \\
    \hline
  \end{tabular}
\end{table}

\begin{figure}
\centering 
\includegraphics*[width=10cm]{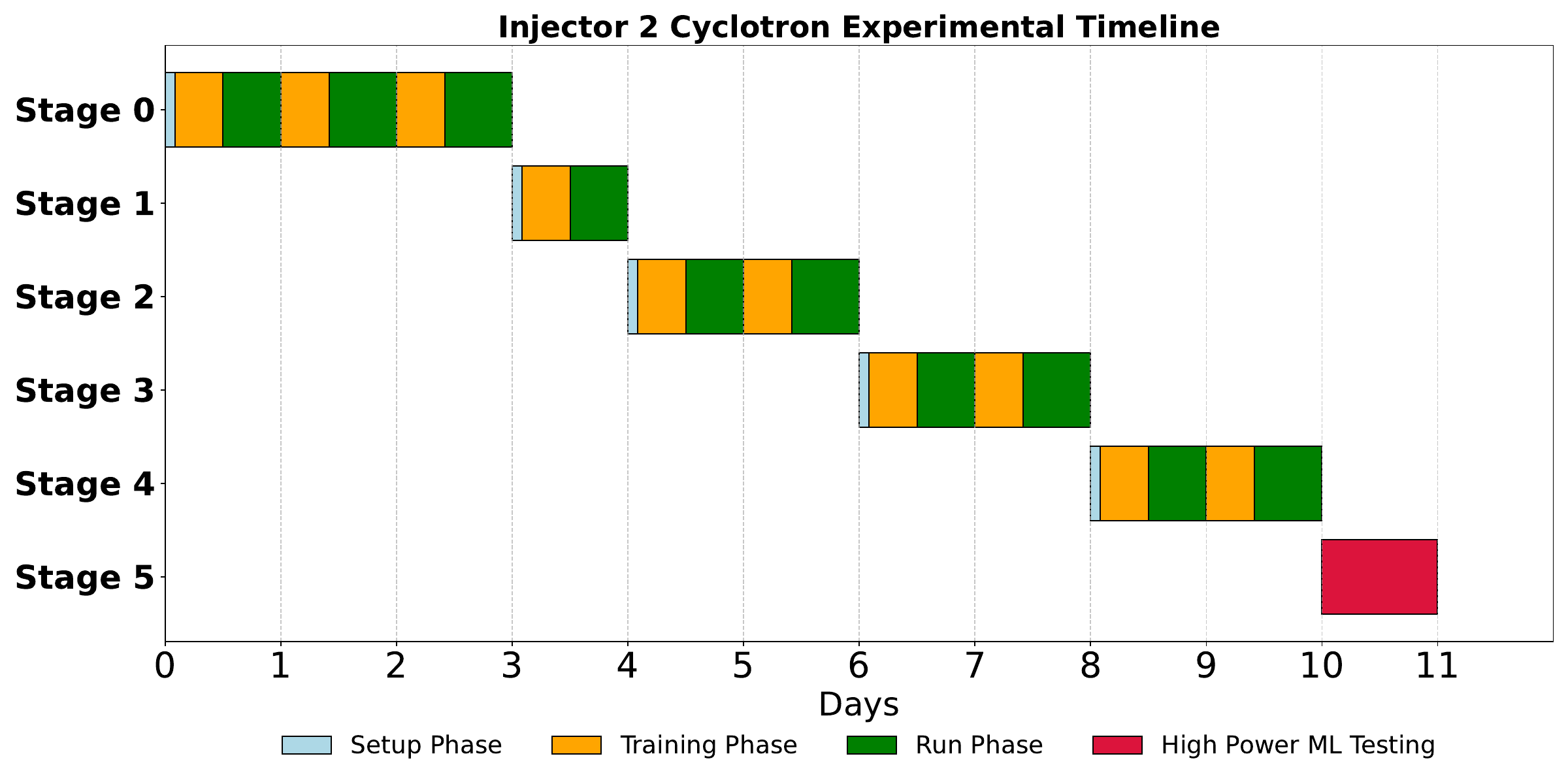}
\caption{Phase Breakdown of Inj 2 Cyclotron Experiment; the experiment took place from Apr 28, 2025 until May~8th, 2025 i.e., for a total of 11 days.}
\label{fig:exp-campaign}
\end{figure}

Each stage of the campaign followed a structured daily cycle. During the morning setup phase, operators configured the machine to the target turn number, optimized the beam at nominal current (2~mA), and probe measurements were performed at various currents. Low-current ML training runs were then performed during the daytime, typically lasting several hours. Following training, the ML agent was left in autonomous control during evening and overnight periods, providing a stringent test of stability, drift compensation, and interlock recovery over many hours of operation (12 hours). A final high-current validation stage 5 extended ML control into the milliampere regime to evaluate the robustness of the learned policies. The overall timeline of the campaign is illustrated in Fig.~\ref{fig:exp-campaign}, which shows the sequential execution of setup, training, and long-term evaluation phases across the 11-day experimental run. An additional one day, prior to experiment, was dedicated for setup tests.
This phased execution strategy ensured that each configuration was benchmarked against operator-validated references, that ML training was conducted in safe regimes, and that robustness was tested under realistic conditions, including day-to-night transitions. In total, the campaign covered five turn numbers across multiple resonator configurations, produced hundreds of hours of machine learning operation, and generated a comprehensive dataset for offline analysis.

\section{Turn and Current-Dependent Beam Dynamics: Radial Profiles and Jacobian Evolution} \label{sec:turn-current-dynamics}
Cyclotron beam dynamics exhibit strong dependence on both the number of turns and the operating beam current. This section summarizes the empirical observations obtained during the Injector~2 campaign, combining radial probe measurements with sensitivity (Jacobian) analyses to quantify how the response of the machine evolves under different conditions.
\subsection{Beam setup and turn-dependent operation}
For each selected turn number configuration, the reference operating point was established at a beam current of 2~mA, which represents the maximum achievable current even in reduced-resonator modes. Daily operations began with low-current operator-led adjustments of resonator voltages and coil settings to configure the desired turn number, followed by optimization at 2~mA to ensure stable transmission and extraction efficiency. In certain cases, additional interventions such as collimator repositioning (KIP4, KIR1L) or adjustment of the extraction septum (AXA) were required to restore beam quality. \\
The complexity of tuning was found to increase with the number of turns, particularly when fewer resonators were active. Configurations with only two active resonators (e.g. 89 turns) exhibited the highest sensitivity, where small deviations in magnetic field or RF voltage led to rapid beam change and frequent interlocks. A summary of the tuning requirements across turn numbers is given in Table~\ref{tab:tuning_summary}, highlighting the longer setup times required in these challenging configurations. Such operator-dependent procedures are not sustainable for ADS-class drivers, thereby motivating the development of ML-based automated tuning methods.
\begin{table}[htb]
  \caption{Summary of operator-led fine-tuning procedures across turn configurations. 
  }
  \label{tab:tuning_summary}
  \centering
  \begin{tabular}{c c c c}
    \hline
    Turn change & Interlocks activated & Tuning time \\
    \hline
    72 $\rightarrow$ 73 & 3 &  10 min \\
    73 $\rightarrow$ 74 & 3 &  9 min \\
    74 $\rightarrow$ 89 & 7 & 47 min \\
    89 $\rightarrow$ 60 & 0 & 10 min \\
    \hline
  \end{tabular}
\end{table}

\subsection{Radial probe measurements}
\begin{figure*}[t]
    \centering
    \subfloat[]{
        \includegraphics[width=0.48\textwidth]{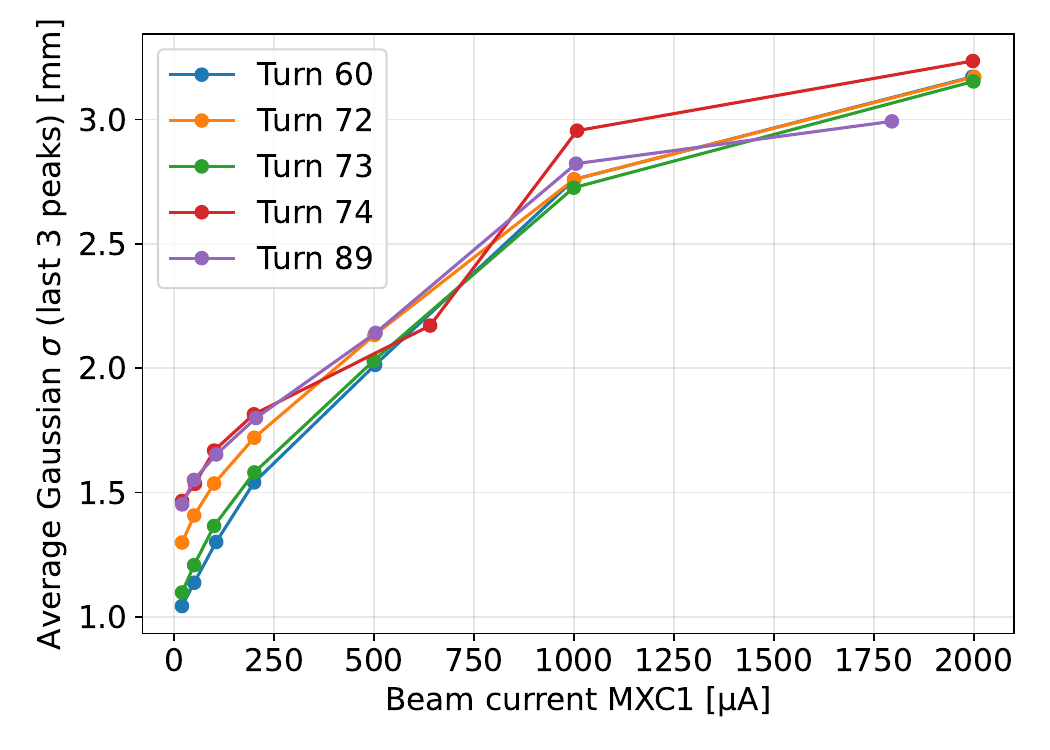}
        \label{fig:radial-3turn}
    }\hfill
    \subfloat[]{
        \includegraphics[width=0.48\textwidth]{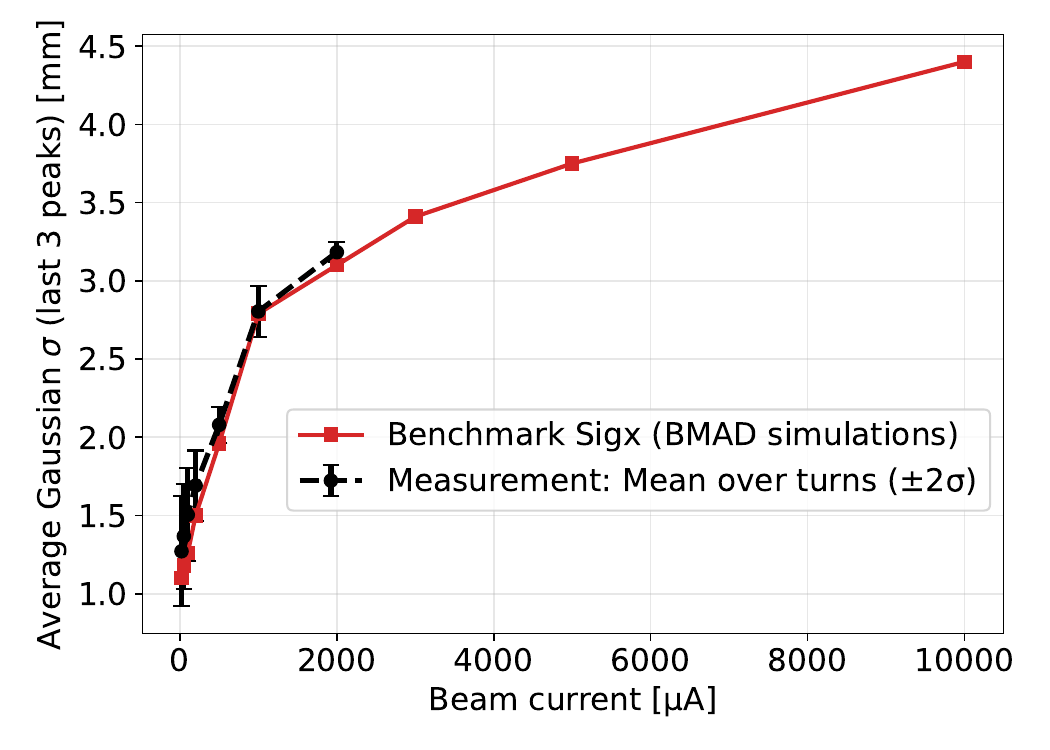}
        \label{fig:radial-sim}
    }
    \caption{(a) Average radial beam size at extraction versus current, 
    for various turn numbers (averaged over the last three turns). 
    (b) Comparison of the measured average $\sigma$, obtained by averaging over 
    all turn numbers shown in (a) (mean $\pm 2\sigma$), with space-charge tracking simulations.}
    \label{fig:radial-scan}
\end{figure*}

To quantify the radial size of the extracted beam, Gaussian fits were applied to the RIE1 probe profiles across beam currents from 20 µA to 2 mA and for multiple turn numbers. The analysis reveals a clear current-dependent broadening of the radial beam size, consistent with earlier observations at injector 2 \cite{BaumgartenZhang, baumgarten2011transverse}, and with systematic differences between turns. To obtain a more representative measure of the extraction region, the radial width was averaged over the last three turns, which smooths fluctuations due to coherent betatron oscillations while preserving the physical envelope. The results, shown in Fig.~\ref{fig:radial-3turn}, demonstrate consistent beam growth with current and turn dependence.

Comparison with BMAD self-consistent space-charge tracking simulations confirms the observed slope change around 1 mA, pointing to the onset of collective effects \cite{PcolBMAD}. The close agreement between experiment and simulation (Fig.~\ref{fig:radial-sim}) provides confidence in the physical interpretation and establishes a solid basis for benchmarking RL-guided tuning strategies under varying current and turn conditions.

\subsection{Machine Nonlinearity and Motivation for RL}
The Jacobian analysis presented in Appendix~\ref{app:jacobian} shows that the beam response varies significantly with turn number and operating point. This strong nonlinearity indicates that a single linear response matrix is insufficient to describe the machine over the investigated operating range, motivating the use of a state-dependent reinforcement learning controller.

Beyond the dependence on operating point, the radial probe and Jacobian measurements indicate that cyclotron tuning becomes progressively more challenging with increasing turn number and beam current. At larger radii, the beam exhibits greater sensitivity to small perturbations, while higher beam currents introduce additional space-charge effects and broaden the beam distribution. Furthermore, slow environmental variations, such as magnet temperature changes, modify the machine response over time. These observations motivated the staged design of the Injector~2 campaign, beginning with low-current training, progressing turn-by-turn, and culminating in validation at higher beam currents. More generally, they highlight the need for adaptive control strategies capable of responding to turn-, current-, and drift-dependent machine behavior rather than relying on a fixed linear response model.

\section{Adapting accelerator physics for ML integration}
The successful deployment of machine learning in accelerator control requires more than sophisticated algorithms; it depends critically on adapting the underlying physics to become accessible, safe, and efficient for data-driven optimization. In Injector 2, a set of targeted modifications were introduced that made the machine’s physical response compatible with RL, effectively bridging the gap between abstract policy learning and practical beam tuning. Two innovations were particularly decisive: an overshooting strategy that accelerated magnet settling times by nearly a factor of six, and a physics-motivated reward design that aligned the agent’s objectives with those of experienced operators.
\subsection{Overshooting Strategy} \label{sec-overshoot}
\begin{figure*}
\centering 
\includegraphics*[width=8cm]{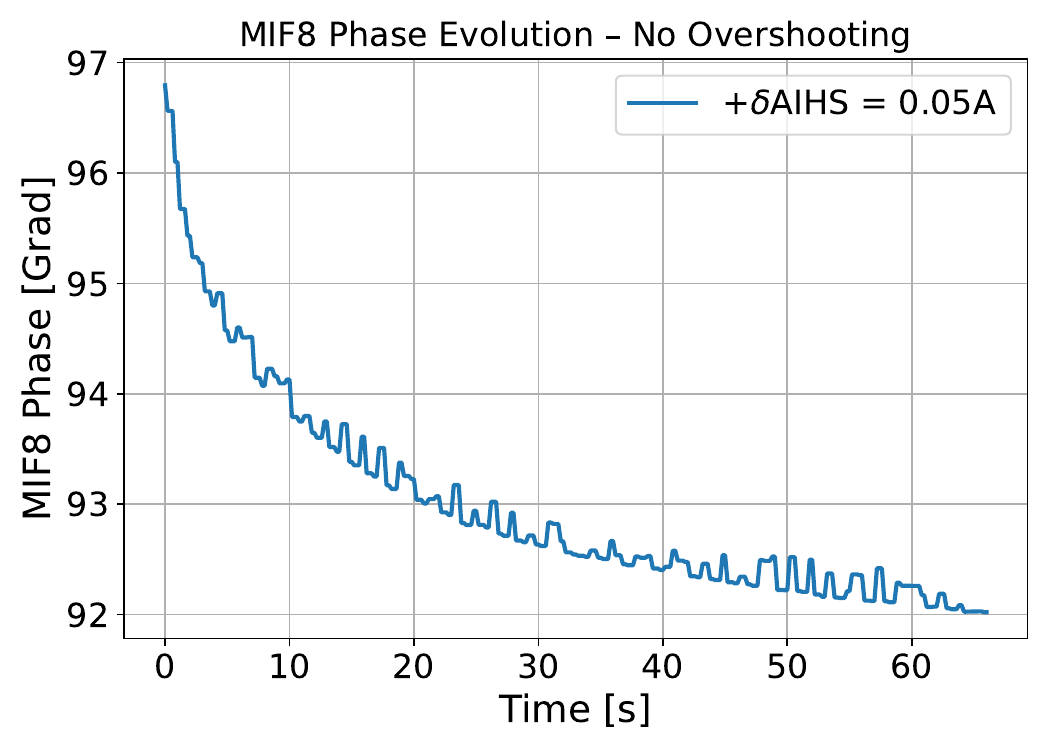}
\caption{MIF8 phase response following a +0.05 A step increase in AIHS current without overshooting. The system exhibits a slow exponential settling, requiring nearly 60 seconds to stabilize (turn 72).}
\label{fig:no_overshooting}
\end{figure*}

One of the primary limitations for online control was the slow settling of the magnetic field following adjustments to the main magnet coil (AIHS). After a small current step, the associated MIF8 phase signal typically required close to 60 s to reach a steady state, as is shown in Fig.~\ref{fig:no_overshooting} severely limiting the number of agent-environment interactions achievable during training. 

\begin{figure*}[t]
    \centering
    \subfloat[]{
        \includegraphics[width=0.48\textwidth]{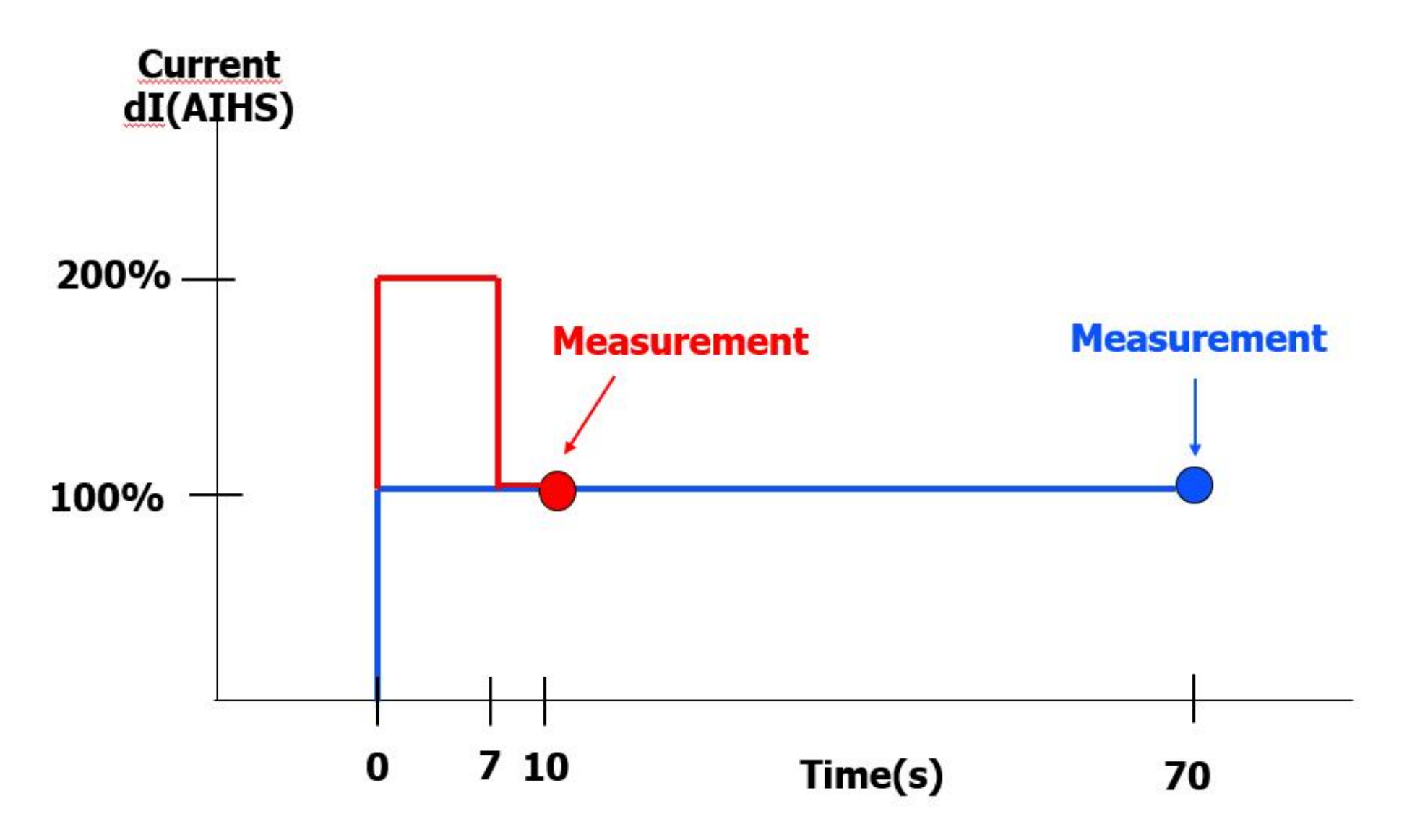}
        \label{fig:overshoot-approach}
    }\hfill
    \subfloat[]{
        \includegraphics[width=0.48\textwidth]{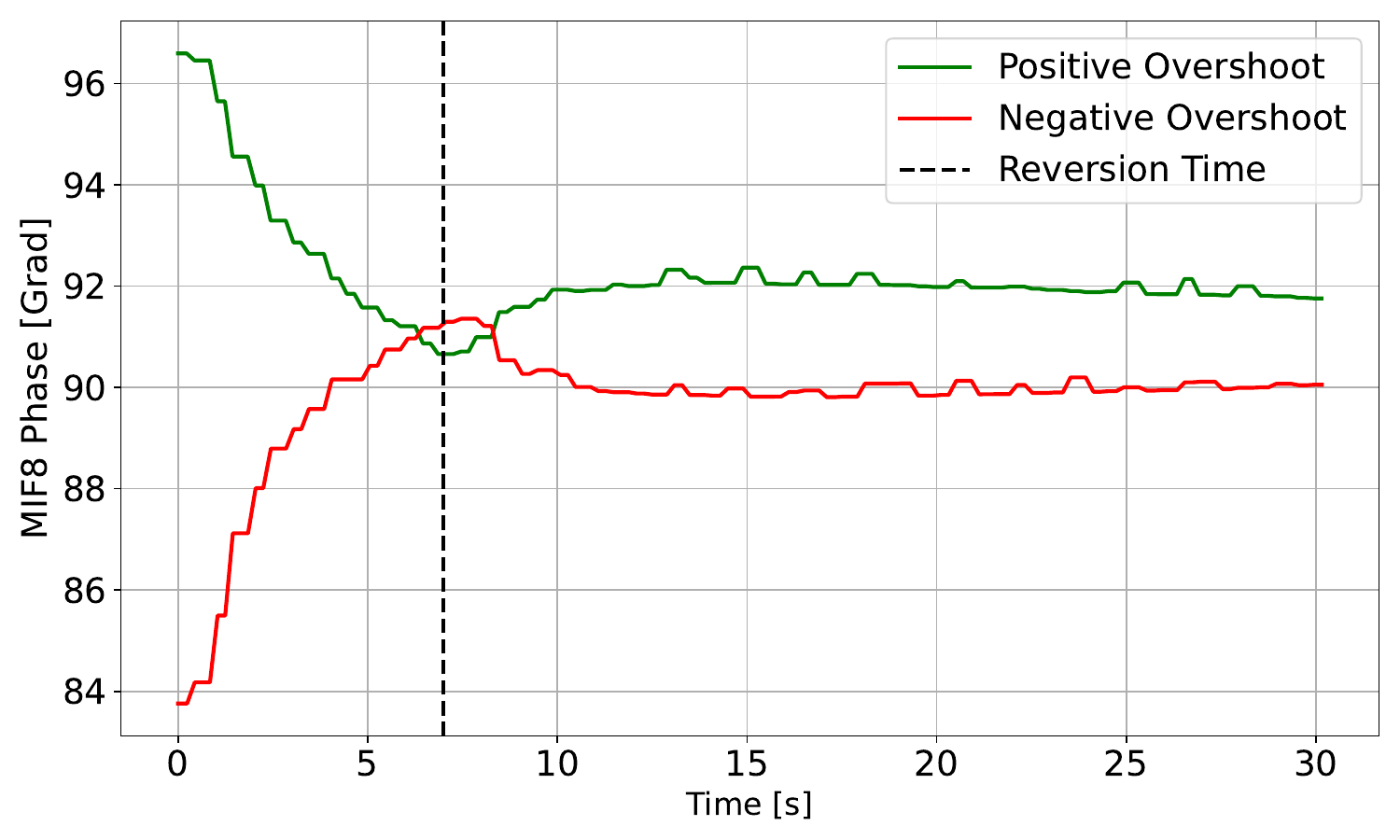}
        \label{fig:overshoot-exp}
    }
    \caption{(a) Main coil current change with (red) or without overshooting (blue). 
    (b) MIF8 phase response to positive and negative overshooting with an AIHS final step of ±0.05 A, using 100\% overshoot and a 7-second reversion time (Turn 72). }
    \label{fig:overshoot}
\end{figure*}

Owing to the slow response of the AIHS main magnet, a simple overshooting strategy was implemented to accelerate magnetic settling. Overshooting is a well-established control technique; here it was adapted to the operational constraints of Injector~2 to reduce the effective settling time from approximately 60 s to 10 s as is shown in Fig.~\ref{fig:overshoot}. This reduction was essential for making online reinforcement-learning experiments practical within the limited beam time available. The method was verified across a range of perturbation amplitudes and constrained to remain within interlock limits.

\subsection{Reward Shaping} \label{sec-reward}
To align the learning objective with accelerator performance and protection, a physics-motivated reward function was designed. The total reward was expressed as:
\begin{eqnarray}
R = R_{beam} + P_{trim} + P_{interlock}
\end{eqnarray}
where $R_{beam}$ encodes beam quality, $P_{trim}$ penalizes excessive use of trim coils, and $P_{interlock}$ enforces safety by strongly penalizing beam loss events.
The central term is the weighted error between measured and reference beam phases 
at the eight MIF probes:
\begin{eqnarray}
R_{\text{phase}} = \left( \sum_{i=1}^{8} w_i \, (\phi_i - \phi_i^{\text{ref}})^2 \right)^{1/2},
\end{eqnarray}
where $\phi_i$ is the measured phase at probe $i$, $\phi_i^{\text{ref}}$ is the corresponding reference, and $w_i$ are weighting factors. The phase weights used throughout the experimental campaign were \[
w=[1,1,1,1,2,3,4,5],
\]
assigning progressively greater importance to the outer probes.
This choice reflects the increasing relevance of phase alignment
near extraction, where beam quality and extraction efficiency are
most sensitive to phase deviations.

Beam losses were incorporated through the deviation of the measured current loss signal $\Delta I_{\text{loss}}$ from its minimal reference value. The combined beam-related reward was then:
\begin{eqnarray}
R_{\text{beam}} = -\tfrac{1}{2} \left[ \tanh\!\left(\frac{R_{\text{phase}}}{10}\right) 
+ \tanh\!\left(\frac{\Delta I_{\text{loss}}}{30}\right) \right] .
\end{eqnarray}
In addition trim-coil usage was penalised to discourage excessive
local corrections:
\begin{eqnarray}
P_{\text{trim}} = -\lambda \frac{1}{N} \sum_{j=1}^{N} |I_j|,
\end{eqnarray}
where $I_j$ is the unnormalised current applied to trim coil $j$, $N=12$ is the number of coils, and $\lambda$ a scaling factor. The latter was selected using the BMAD simulation environment prior to deployment on the real machine.
Such an approach ensures bounded contributions and avoids divergence during training.

The reward function was first implemented in a BMAD-based tracking environment, which provided a safe platform for systematic testing. These simulations confirmed that the chosen formulation was sensitive to physically meaningful deviations and yielded stable agent behavior. In particular, they allowed inspection of the RL agent’s convergence time under different reward designs. Good convergence was defined as achieving successful tuning within fewer than 1000 timesteps, typically corresponding to the agent bringing the machine to the desired state in one or two corrective actions. The final reward formulation consistently met this criterion, ensuring both fast convergence and physically interpretable actions. This validation step was essential prior to deployment on the real machine.

The reward threshold of -0.08 was selected as a compromise between the intrinsic noise level of the phase diagnostics and the accelerator physics definition of successful beam tuning. This value was fine-tuned using simulation studies to ensure consistent convergence behavior. In practice, the corresponding performance, phase errors of approximately ±1° and average trim-coil usage below 20\% of their range, represents a physically meaningful criterion for optimal tuning equivalent to $R >-0.08$.

\subsection{Interlock Handling}
A layered interlock strategy was implemented in addition to the standard HIPA protection systems. 
The interlocks monitored three classes of signals: (i) the validity of the MIF phase measurements, 
(ii) the level of beam losses detected by KXAI, MII7, and MXI1, and (iii) the stability of the extracted current measured by MXC1. 
An interlock was issued whenever the MIF signals became invalid or out of range, whenever losses exceeded the predefined thresholds or returned undefined values, or whenever the MXC1 current deviated by more than 20\% from its nominal value.  

The interlock response followed a two-stage logic. In the first stage, the ML agent rolled back to the last known valid action with a small perturbation added to avoid repeated oscillation around the same point. 
If recovery was not achieved within the allowed window, a second interlock was raised. In this case, RL control was halted and the system transitioned into a safe mode, in which it remained until the MXC1 signal demonstrated stable behavior within 20\% for at least 40~s.  

All interlock events were automatically logged together with their timestamps, type (single or double), and root cause, which allowed a detailed analysis of agent behavior under fault conditions.

\section{Reinforcement Learning Methodology}
The reinforcement learning framework was implemented using the Twin Delayed Deep Deterministic Policy Gradient (TD3) algorithm, which is well suited for continuous action spaces and has been shown to provide stable convergence in noisy environments \cite{Fujimoto2018TD3, Kain.23.124801}. A custom OpenAI Gym–compatible environment was developed to interface the agent with Injector~2, ensuring modularity and reproducibility of the experimental framework.

\begin{table}[htb]
\caption{Main TD3 hyperparameters used during the Injector~2 campaign.}
\label{tab:td3_hyperparameters}
\centering
\begin{tabular}{lc}
\hline
Parameter & Value \\
\hline
Actor network & [256,256] ReLU \\
Critic network & [256,256] ReLU \\
Learning rate & $10^{-3}$ \\
Batch size & 64 \\
Replay buffer size & $10^6$ \\
Learning Starts & $50$ \\
Discount factor $\gamma$ & 0.99 \\
Target update coefficient $\tau$ & 0.005 \\
Policy update delay & 2 \\
Target policy noise & 0.2 \\
Target noise clip & 0.5 \\
Maximum episode length & 40 steps \\
Action space & 14 continuous variables \\
Observation space & 14 continuous variables \\
\hline
\end{tabular}
\end{table}

Injector~2 is operated with the Experimental Physics and Industrial Control System (EPICS). A dedicated Linux server running PyEPICS \cite{pyepics} provided real-time access to process variables (PVs), enabling safe, low-latency communication between the ML agent and the accelerator subsystems.

The observation vector consisted of 14 continuous variables:
the eight phase measurements MIF1--MIF8,
the two loss monitors MII7 and MXI1,
the extracted beam current MXC1,
the estimated turn number,
the mean magnet temperature MIT\_M,
and the bunker temperature MIT.
Together these quantities provide information on beam quality,
losses, operating point, and slow machine drifts.

The action space also consisted of 14 continuous variables:
the twelve trim-coil currents (TI1A/B--TI11),
the main correction coil current AIHS,
and the resonator voltage CI3V.
All state and action variables were normalized to the interval
[-1,1] prior to training.

The agent’s objective was to minimize the weighted deviation between the measured phase profile and a predefined reference, while simultaneously penalizing beam losses and excessive trim coil usage (see Sec. \ref{sec-reward}). Interlock conditions defined by abrupt beam current drops, anomalous losses, or invalid phase signals, resulted in large negative rewards, enforcing safe exploration during training.

The slow response of the main coil (AIHS) was addressed through a dedicated overshooting strategy (Sec. \ref{sec-overshoot}), which reduced magnetic settling times from approximately 60 s to 10 s. As a result, each action-observation cycle was executed on a fixed 10 s timescale, setting the minimum step duration during training episodes.

The training of the RL agent was organized into episodes, each corresponding to an independent tuning attempt starting from a given machine configuration. The maximum episode length was defined as the maximum number of agent-environment interactions allowed per attempt, and represents a key hyperparameter for balancing exploration and training efficiency. Each interaction consisted of proposing a new action, applying it to the machine via PyEPICS, allowing the system to stabilize (using the overshooting strategy for the main coil), and recording the resulting observations and reward. Although each interaction required approximately 10~s once an episode was underway, the total wall-clock training time also included the reset procedure performed at the beginning of every episode. This reset typically required about 40~s to restore the machine to a valid initial condition and could occasionally take longer if an interlock was triggered during the reset process. Consequently, the total training duration reported in this work corresponds to the measured wall-clock time, including both RL interaction steps and episode reset overheads, rather than simply the number of timesteps multiplied by 10~s.
For Injector~2 training, the maximum episode length was set to 40 steps, which was sufficient to allow the agent to explore corrective actions while keeping training cycles short and compatible with operational constraints. Episodes terminated either when the maximum number of steps was reached or when the cumulative reward surpassed a predefined threshold indicating successful tuning. The principal TD3 hyperparameters used during the Injector~2 campaign are summarized in Table~\ref{tab:td3_hyperparameters}. Most values correspond to the default implementation provided by Stable-Baselines3 \cite{raffin2021stable}. Although limited studies were performed on learning rates, batch sizes, and exploration noise levels, no significant performance improvements were observed relative to the default configuration.

In contrast, the convergence speed, stability, and safety of the RL agent were found to depend much more strongly on the design of the reward function and the incorporation of accelerator-specific constraints. In particular, reward shaping, interlock penalties, action-space limitations, and the overshooting strategy had a substantially larger impact on training performance than moderate variations of the TD3 hyperparameters. Consequently, the focus of the optimization effort was placed on the formulation of the control problem rather than extensive hyperparameter tuning.

To investigate how prior knowledge influences convergence, three complementary training strategies were tested (Fig.~\ref{fig:rl_strategies}):
(a) RL from scratch, where the agent was initialized with random weights;
(b) RL from previous turn, in which the policy learned at one turn number was reused as initialization for the next; and
(c) RL from surrogate model, where the actor network was pretrained offline at selected turns using a surrogate derived from historical Injector~2 data, before fine-tuning directly on the machine.

Together, these approaches enabled a systematic assessment of learning performance, policy transfer, and the value of physics-inspired pretraining.

\begin{figure*}[t]
    \centering
    \begin{tikzpicture}[node distance=1.8cm and 2.2cm, >=Stealth]

        \node[draw, rounded corners, thick, align=center, minimum width=3.5cm, minimum height=1.2cm, fill=blue!10] (scratch) {RL from Scratch \\ \footnotesize Random initialization};
        
        \node[draw, rounded corners, thick, align=center, minimum width=3.5cm, minimum height=1.2cm, fill=green!10, right=of scratch] (transfer) {RL from Previous Turn \\ \footnotesize Policy transfer across turns};

        \node[draw, rounded corners, thick, align=center, minimum width=3.5cm, minimum height=1.2cm, fill=orange!10, right=of transfer] (surrogate) {RL from Surrogate Model \\ \footnotesize Pretraining on historical data};

        \node[draw, rounded corners, thick, align=center, minimum width=12cm, minimum height=1.2cm, below=3cm of transfer, fill=gray!10] (inj2) {Injector~2 Environment \\ \footnotesize \quad\textbullet\quad Real-time diagnostics \& control via PyEPICS, \quad\textbullet\quad 10\,s step via overshoot};

        \draw[->, thick] (scratch) -- (inj2);
        \draw[->, thick] (transfer) -- (inj2);
        \draw[->, thick] (surrogate) -- (inj2);

    \end{tikzpicture}
    \caption{Overview of the three RL training strategies explored in the Injector~2 experiment: (a) training from scratch with no prior knowledge, (b) initializing from a policy trained at a neighboring turn number, and (c) pretraining with a surrogate model derived from historical data before online fine-tuning.}
    \label{fig:rl_strategies}
\end{figure*}

\section{Experimental Results}
We evaluated the TD3 agent across five turn configurations spanning nominal and degraded operating modes. Performance was assessed via: (i) convergence time (timesteps and wall-clock), (ii) episode length at convergence (target $\leq$ 10 steps), (iii) final weighted phase error (near-extraction emphasis), (iv) beam losses relative to warning thresholds, and (v) number of interlocks (safety). A consolidated view is provided in Table~\ref{tab:rl_summary}.
\begin{table}[t]
\caption{Summary of TD3 training performance across turns. Convergence time is reported in steps (wall-clock hours).}
\label{tab:rl_summary}
\centering
\begin{ruledtabular}
\begin{tabular}{lcccccc}
Turn & Resonators Active & Pretraining & Convergence & Avg.\ Final Reward & Interlocks & Notable Outcome \\
\hline
72 & 3 (R1,R2,R3) & No  & 535 (\,$\sim$4.0 h)   & $-0.06$   & 21 (3 post-conv) & From scratch \\
73 & 3 (R1,R2,R3) & Yes & 291 (\,$\sim$2.3 h) & $-0.06$   & 12 (0 post-conv) & Surrogate pretraining \\
74 & 3 (R1,R2,R3) & Yes & 1117 (\,$\sim$5.8 h)& $-0.06$   & 51 (3 post-conv) & Transfer from Turn 73 \\
89 & 2 (R1,R3)    & Yes & 114 (\,$\sim$0.9 h) & $-0.055$  & 0                & Degraded config; successful \\
60 & 4 (R1–R4)    & No  & 217 (\,$\sim$2.1 h) & $-0.06$   & 4 (1 post-conv)  & Nominal config \\
\end{tabular}
\end{ruledtabular}

\vspace{0.3em}
\footnotesize Convergence is defined as the point after which all episodes solve in $\leq$\,10 steps.
\end{table}

\subsection{Turn-by-turn RL Training Performance}
The learning curves presented in this section characterize the online adaptation of the TD3 agent under different initialization strategies, including training from scratch, surrogate-based pretraining, and transfer from a previously trained turn configuration. They should therefore not be interpreted as a direct comparison with experienced operators or conventional feedback systems. In particular, operators rely on substantial prior machine experience, while response-matrix-based feedback requires prior system identification and is typically valid only near a fixed working point. The purpose of these curves is instead to quantify how rapidly each initialization strategy produces a usable state-dependent control policy on the real machine. The effectiveness of the resulting policies is then assessed through beam-loss reduction, phase alignment, interlock behavior, and subsequent autonomous evaluation.

\begin{figure}[t]
\centering 
\includegraphics*[width=8cm]{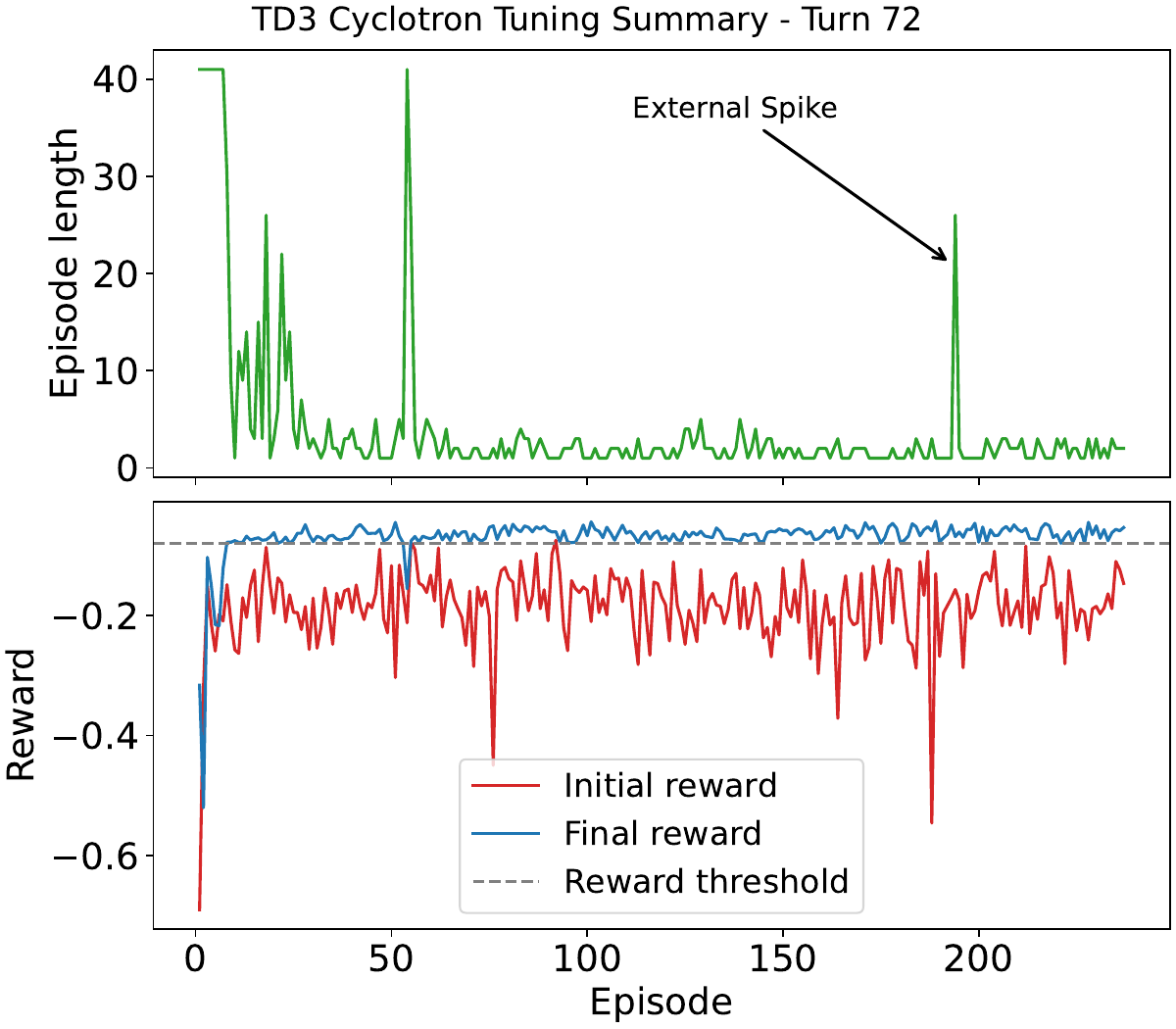}
\caption{Learning Curve for Turn 72: Top Plot, Episode Length during TD3 Training; bottom Plot, Reward evolution across episodes. Total training time is 960 timesteps, totaling 7 hours from start-to-end.}
\label{fig:reward-turn72}
\end{figure}

\begin{figure*}[t]
    \centering
    \begin{minipage}[t]{0.48\textwidth}
        \centering
        \includegraphics[width=0.8\linewidth]{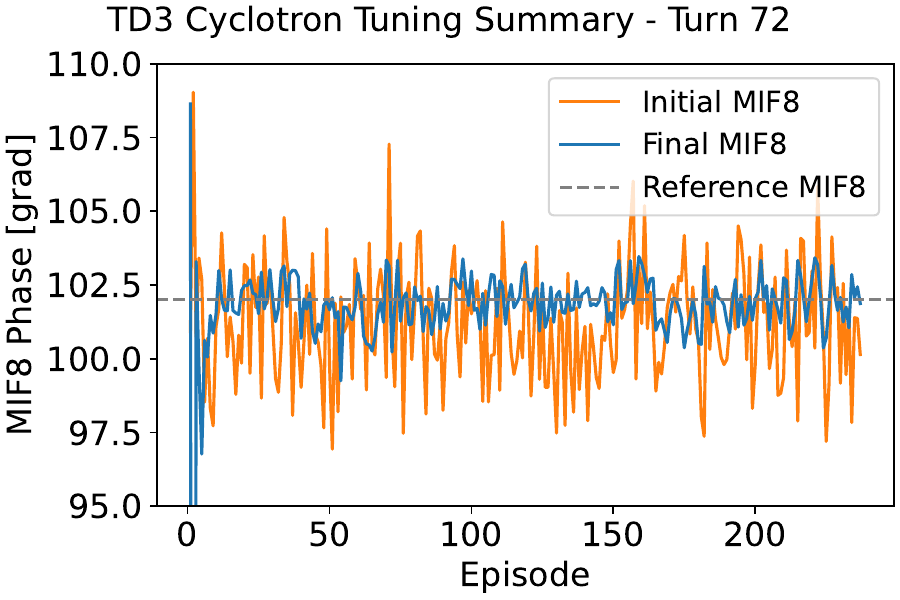}\\[0.8em]
        \includegraphics[width=0.8\linewidth]{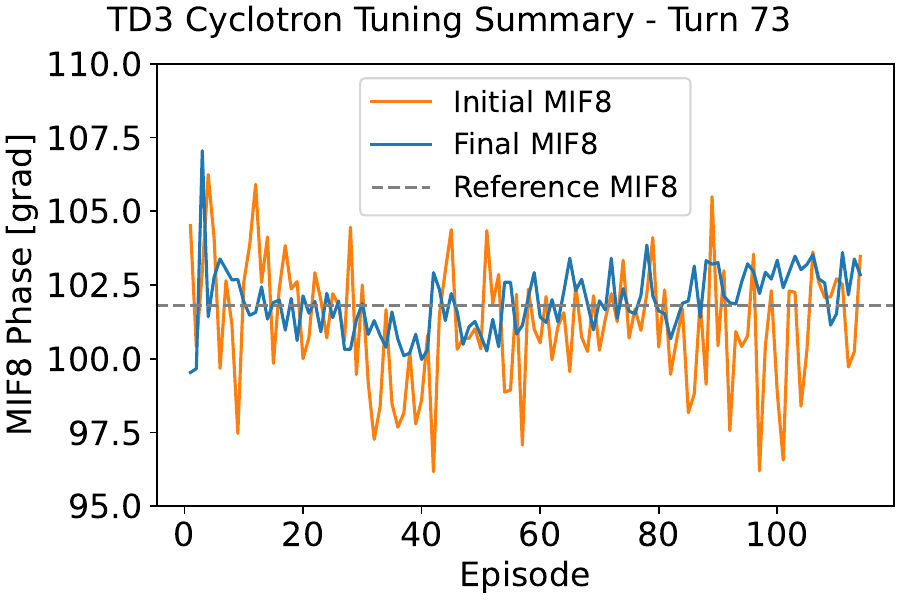}\\[0.8em]
        \includegraphics[width=0.8\linewidth]{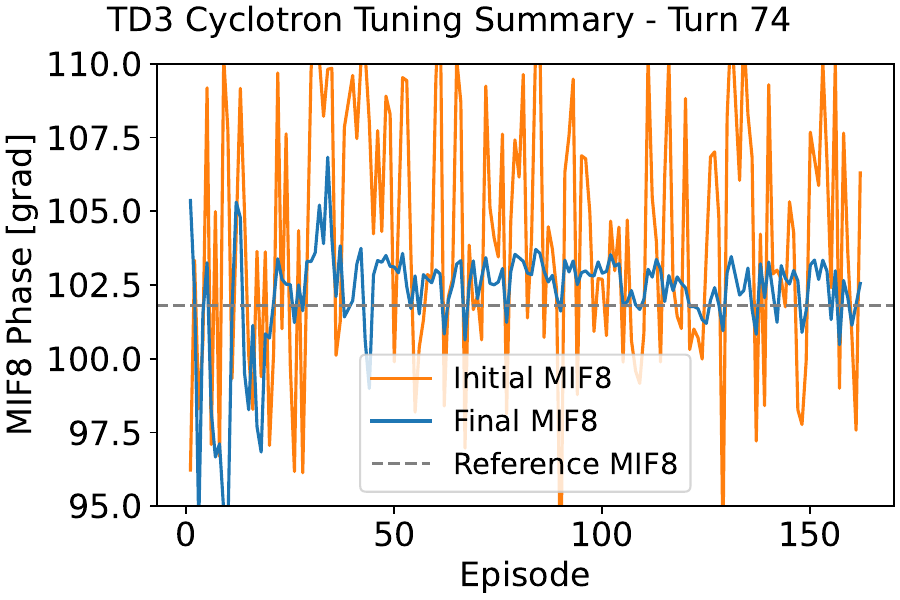}\\[0.8em]
        \includegraphics[width=0.8\linewidth]{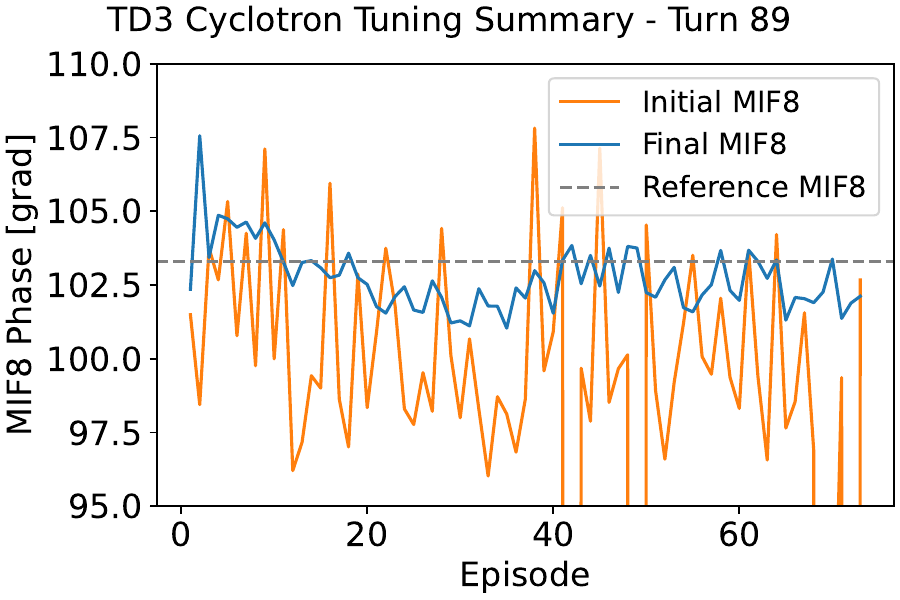}\\[0.8em]
        \includegraphics[width=0.8\linewidth]{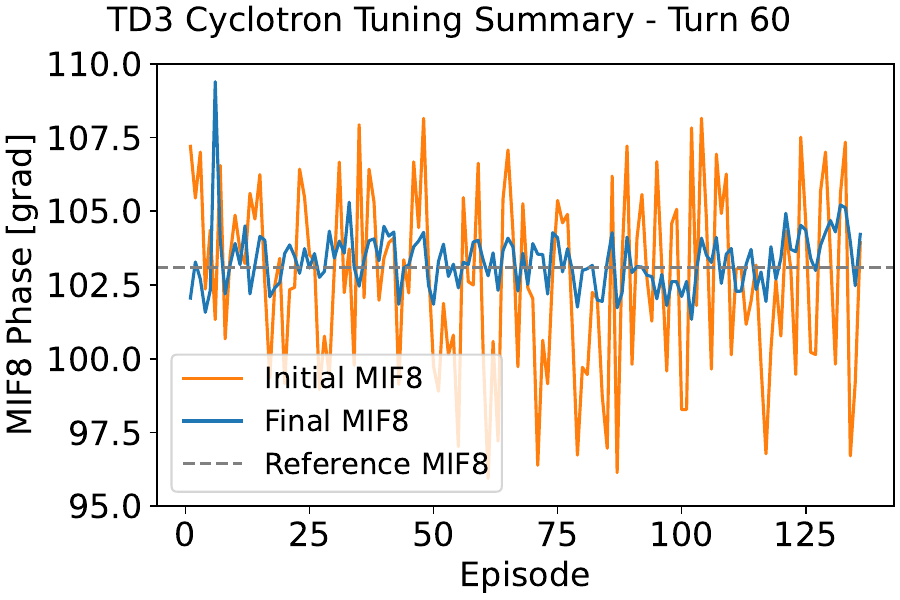}
    \end{minipage}\hfill
    \begin{minipage}[t]{0.48\textwidth}
        \centering
        \includegraphics[width=0.8\linewidth]{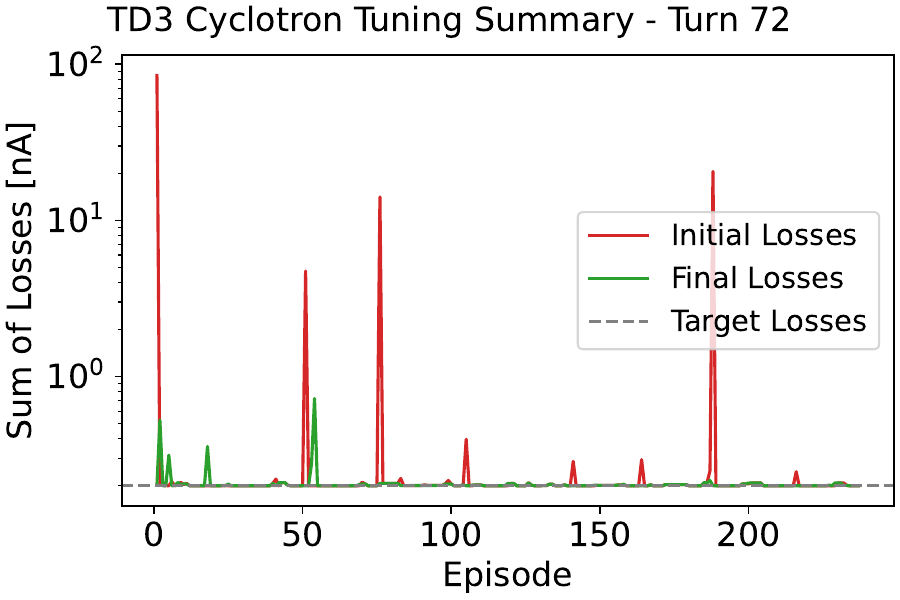}\\[0.8em]
        \includegraphics[width=0.8\linewidth]{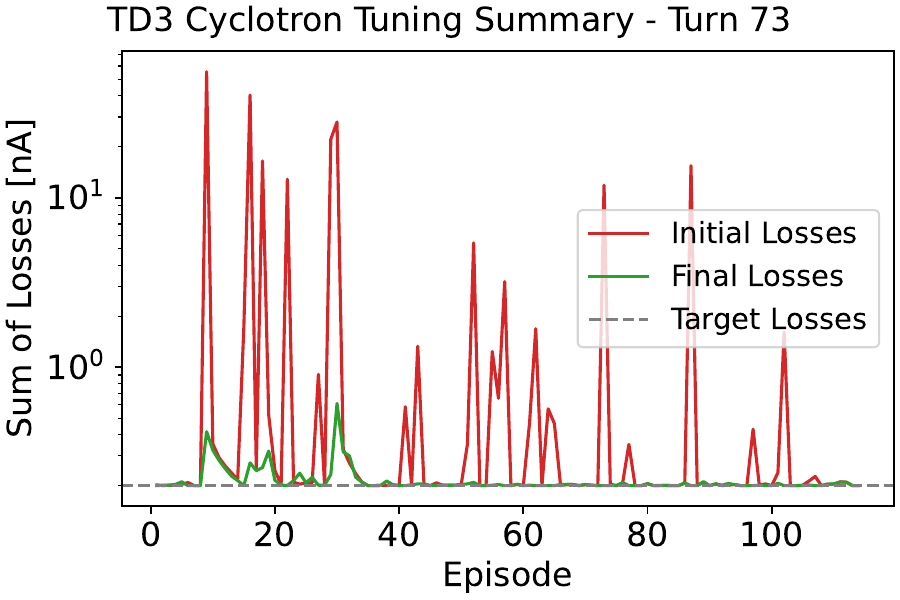}\\[0.8em]
        \includegraphics[width=0.8\linewidth]{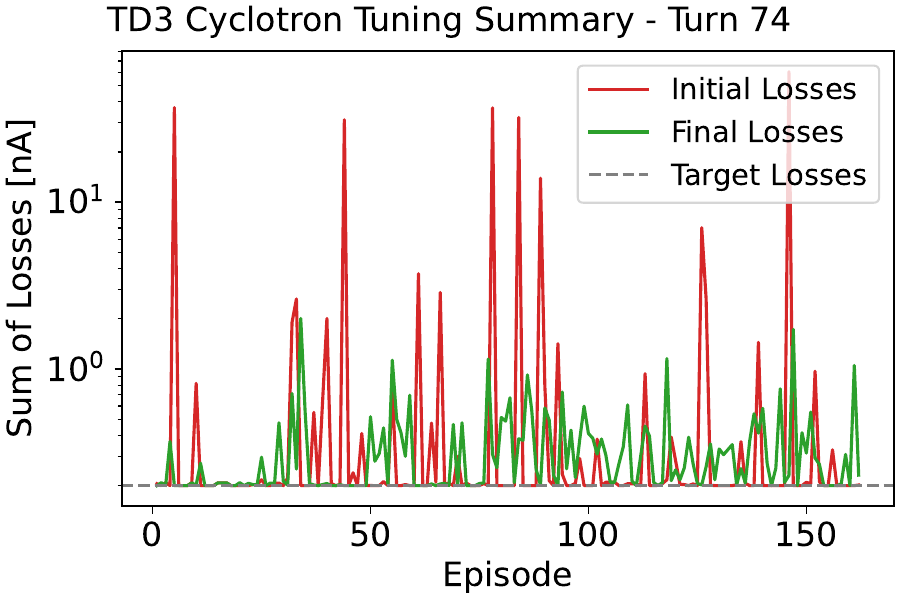}\\[0.8em]
        \includegraphics[width=0.8\linewidth]{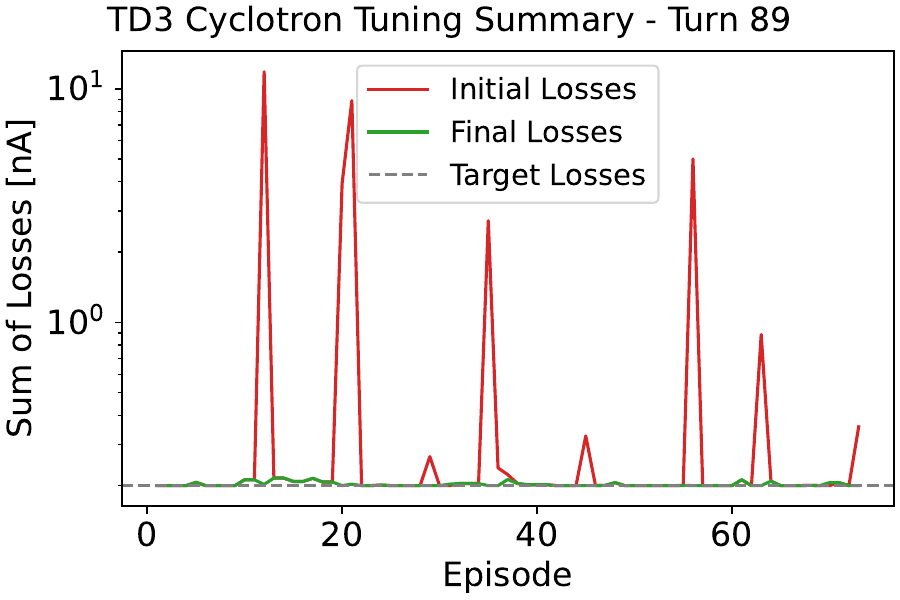}\\[0.8em]
        \includegraphics[width=0.8\linewidth]{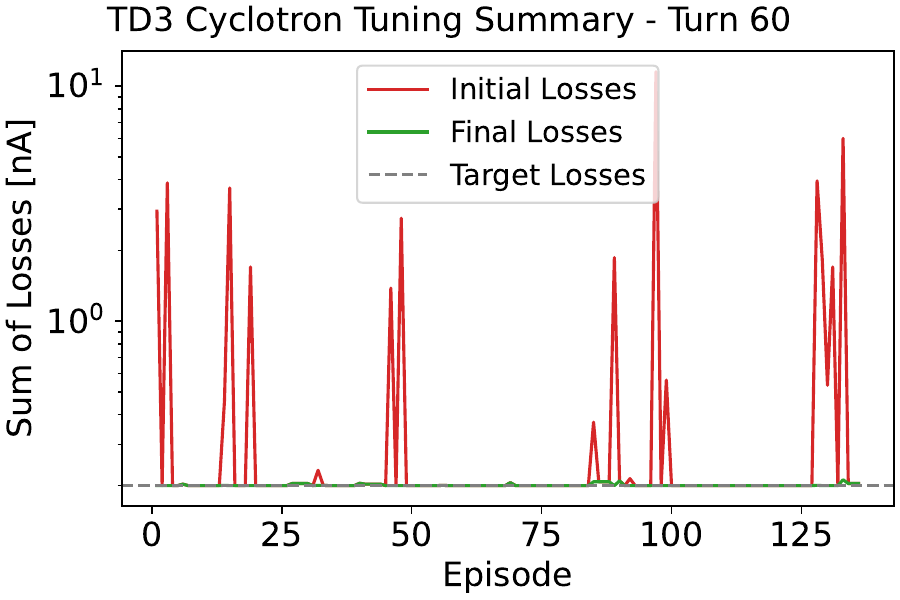}
    \end{minipage}

    \caption{Summary across five turn configurations (top to bottom: 72, 73, 74, 89, 60). 
    Left column: phase-alignment performance (MIF8 at extraction) versus training episode. 
    Right column: beam losses at extraction versus training episode. 
    Initial denotes the machine state before the agent's first action. Final represents the last state reached by the RL agent at the end of the episode. }
    \label{fig:phase-loss-stacks}
\end{figure*}

\begin{figure*}[t]
    \centering
    \begin{minipage}[t]{0.48\textwidth}
        \centering
        \includegraphics[width=0.8\linewidth]{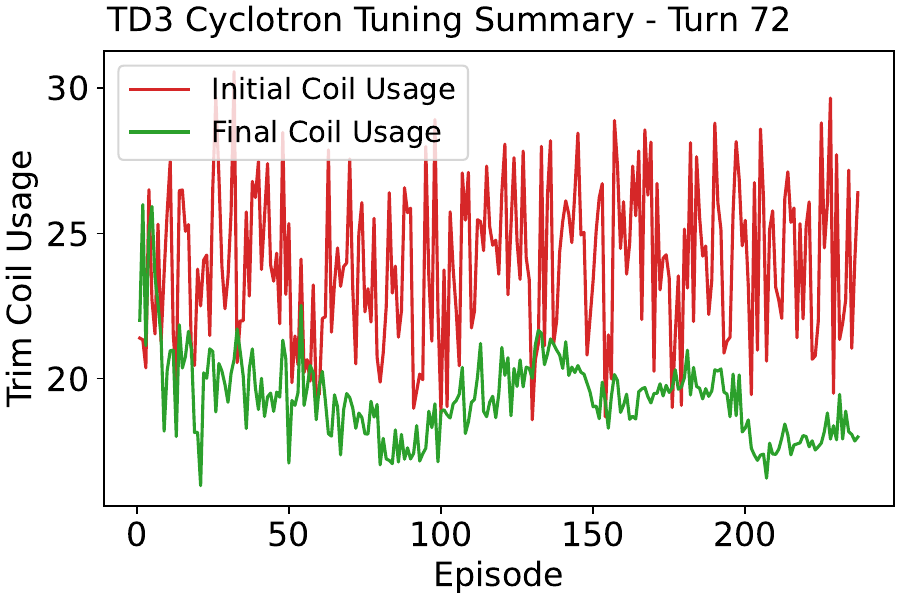}\\[0.8em]
        \includegraphics[width=0.8\linewidth]{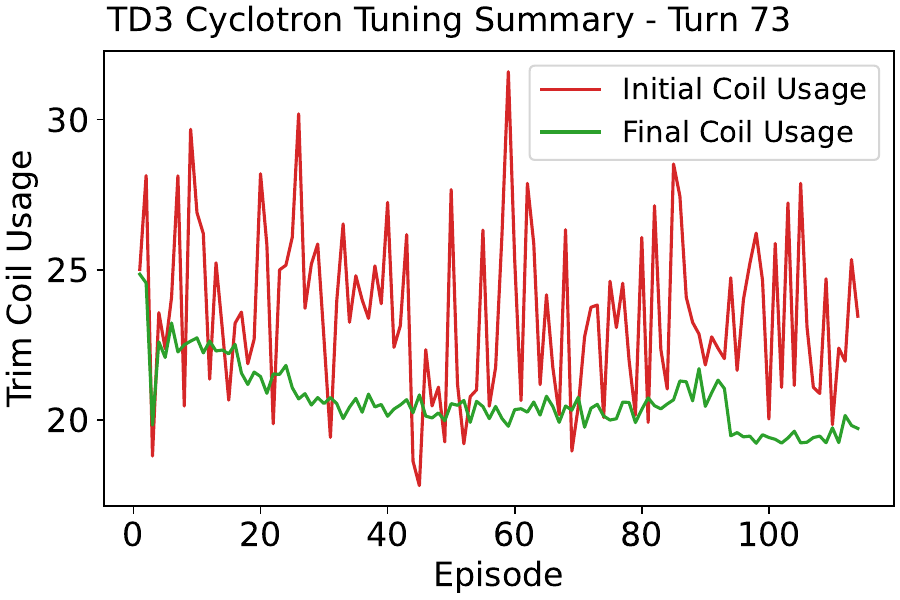}\\[0.8em]
        \includegraphics[width=0.8\linewidth]{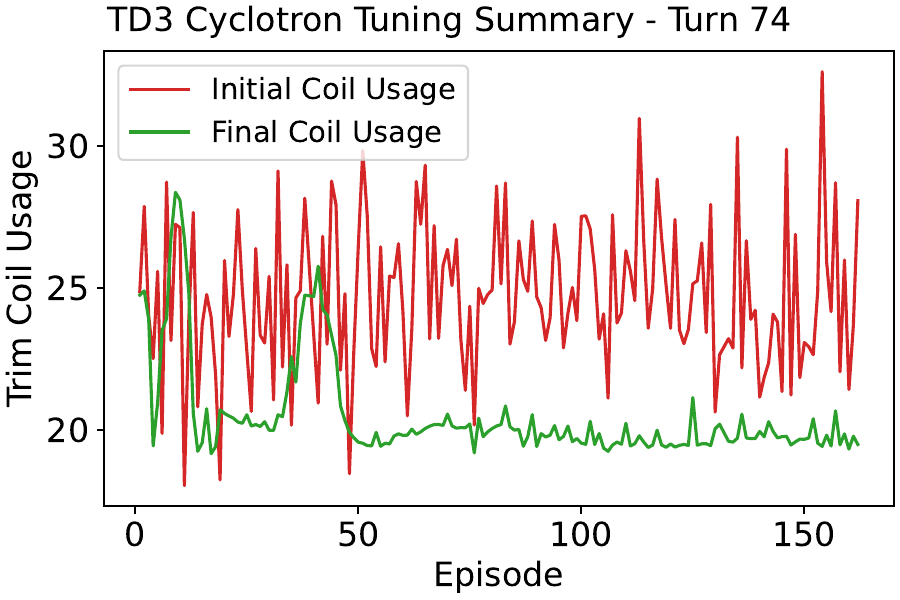}\\[0.8em]
        \includegraphics[width=0.8\linewidth]{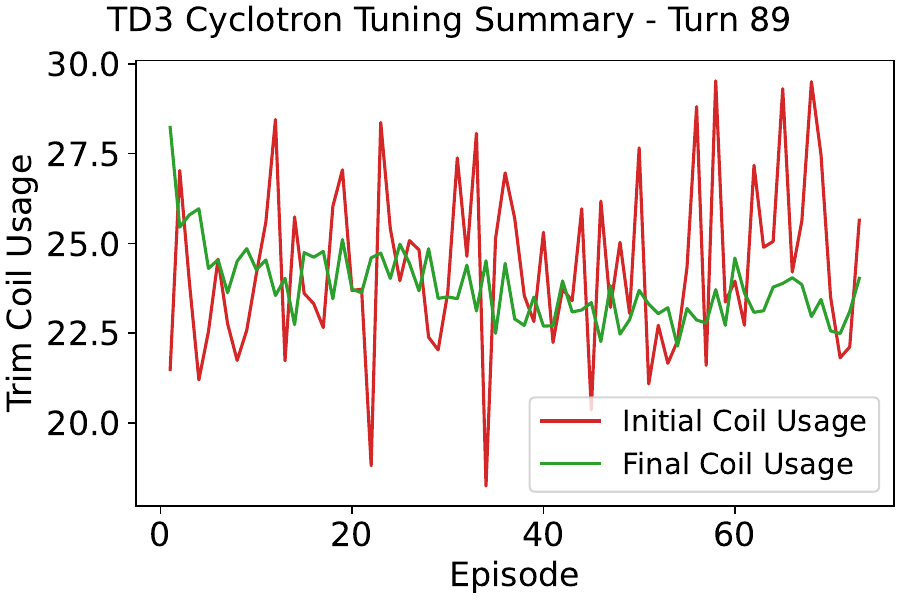}\\[0.8em]
        \includegraphics[width=0.8\linewidth]{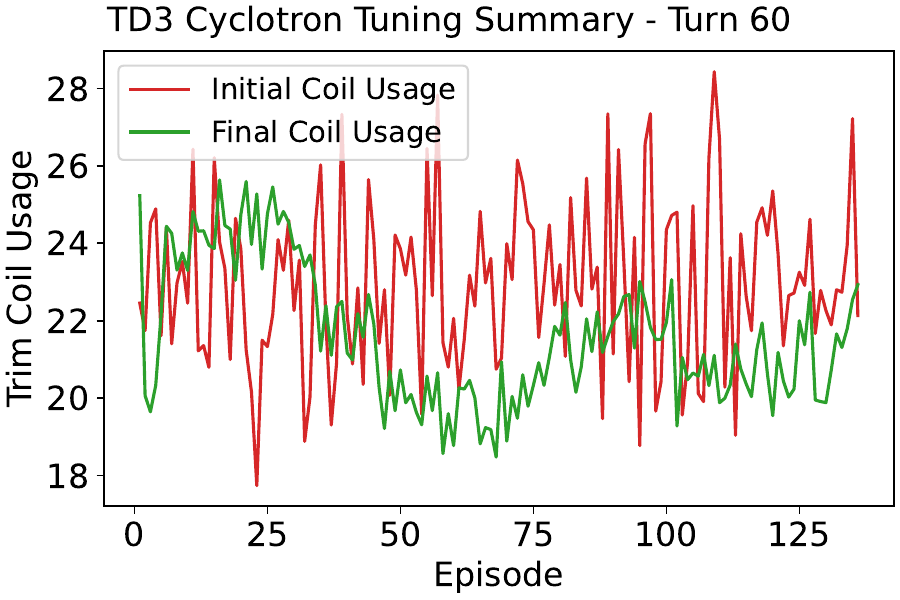}
    \end{minipage}\hfill
    \begin{minipage}[t]{0.48\textwidth}
        \centering
        \includegraphics[width=0.8\linewidth]{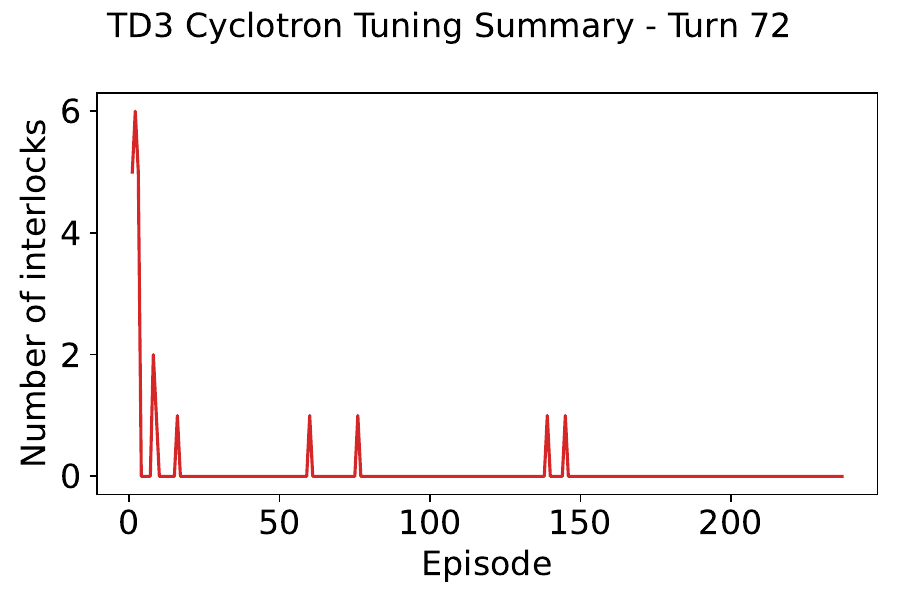}\\[0.8em]
        \includegraphics[width=0.8\linewidth]{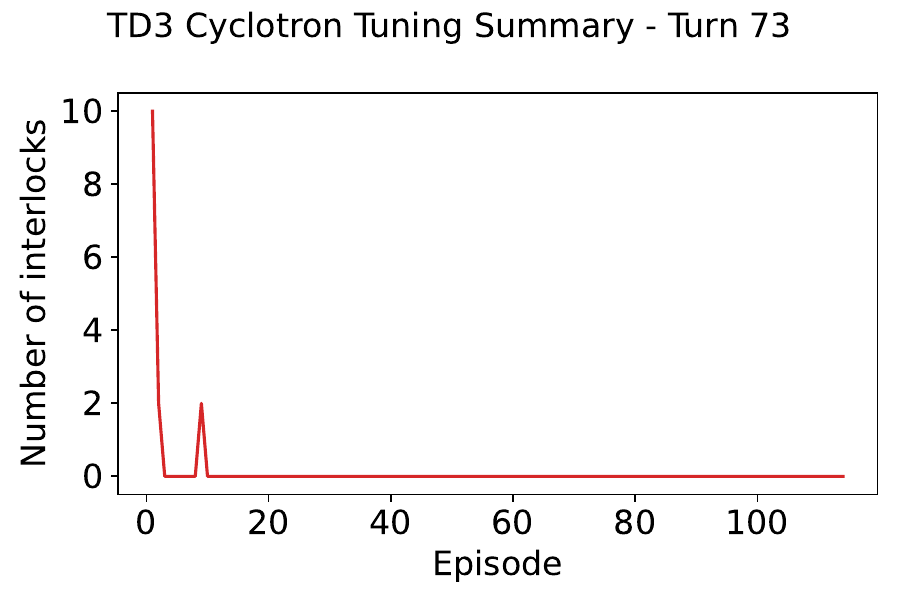}\\[0.8em]
        \includegraphics[width=0.8\linewidth]{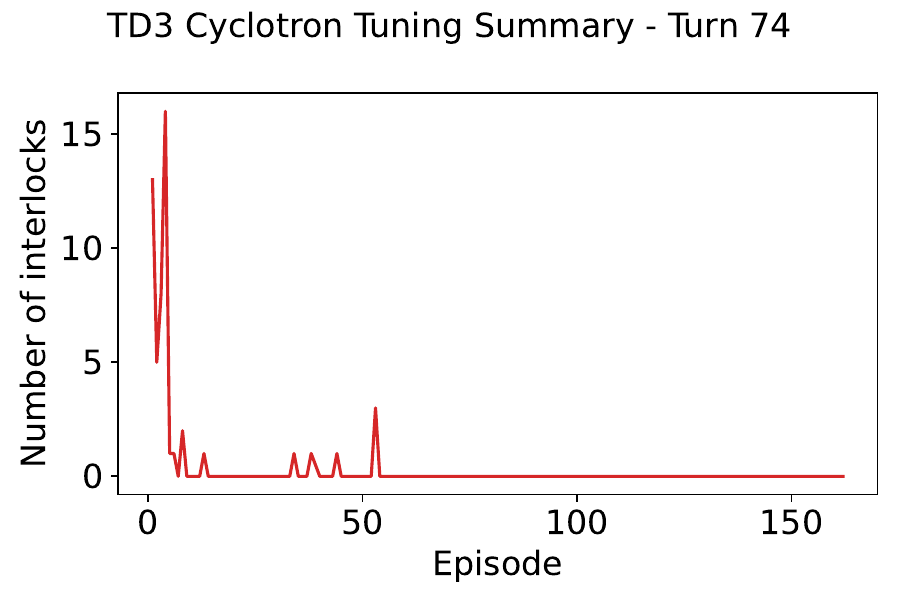}\\[0.8em]
        \includegraphics[width=0.8\linewidth]{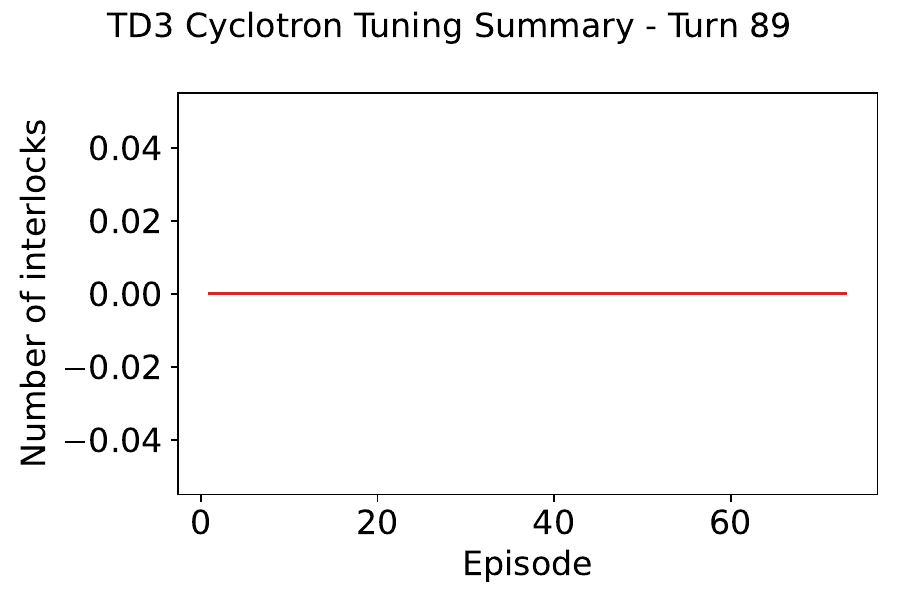}\\[0.8em]
        \includegraphics[width=0.8\linewidth]{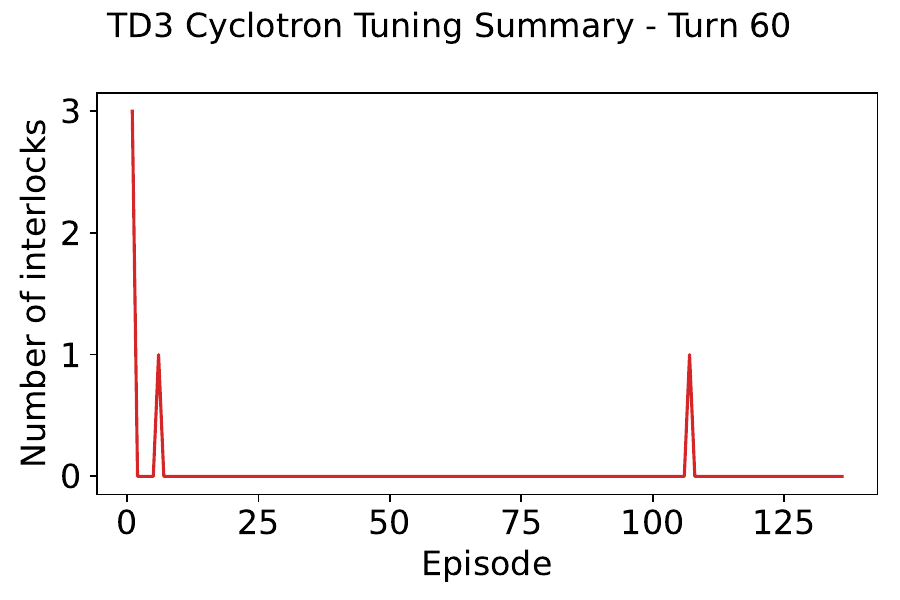}
    \end{minipage}

    \caption{Summary across five turn configurations (top to bottom: 72, 73, 74, 89, 60). 
    Left column: coil usage versus training episode. 
    Right column: number of interlocks versus training episode. 
    All cases converge to the operational targets; see text for per-turn details.}
    \label{fig:coil-interlocks-stacks}
\end{figure*}

\textbf{Turn 72 (baseline, from scratch):}  
This case established the reference performance of the RL agent without any prior information. Starting from random initialization, the agent required about 4 hours (535 steps) to converge. The learning curve in Fig.~\ref{fig:reward-turn72} shows the drop of episode length to $\leq 10$ steps with stable final returns. Phase error collapses to within $\pm 1$ degree (Fig.~\ref{fig:phase-loss-stacks}, left column, row~1), losses fall well below warning thresholds (right column, row~1), and interlocks are confined to early exploration (Fig.~\ref{fig:coil-interlocks-stacks}, right column, row~1). A total of 21 interlocks were triggered during this training, decreasing to only 3 after convergence.

Beam losses and coil efficiency improved markedly over training. As seen in Fig.~\ref{fig:phase-loss-stacks} (right column, row 1), losses decreased by nearly two orders of magnitude, reaching levels far below the 30 nA warning threshold. In parallel, the mean absolute trim-coil currents were reduced (Fig.~\ref{fig:coil-interlocks-stacks}, left column, row 1), indicating that the agent converged to efficient control strategies that achieve precise phase alignment with minimal corrective action.

\textbf{Turn 73 (comparison: surrogate pretraining vs. from scratch):}  
At Turn 73, two RL runs were performed under identical machine conditions to directly assess the impact of pretraining.
In the first case, the agent was initialized from scratch (random weights), while in the second, the same agent was initialized using a surrogate model pretrained on historical Injector 2 data. This setup enabled a direct, controlled comparison between pretrained and non-pretrained policies for the same turn number.

\begin{figure*}
\centering 
\includegraphics*[width=8cm]{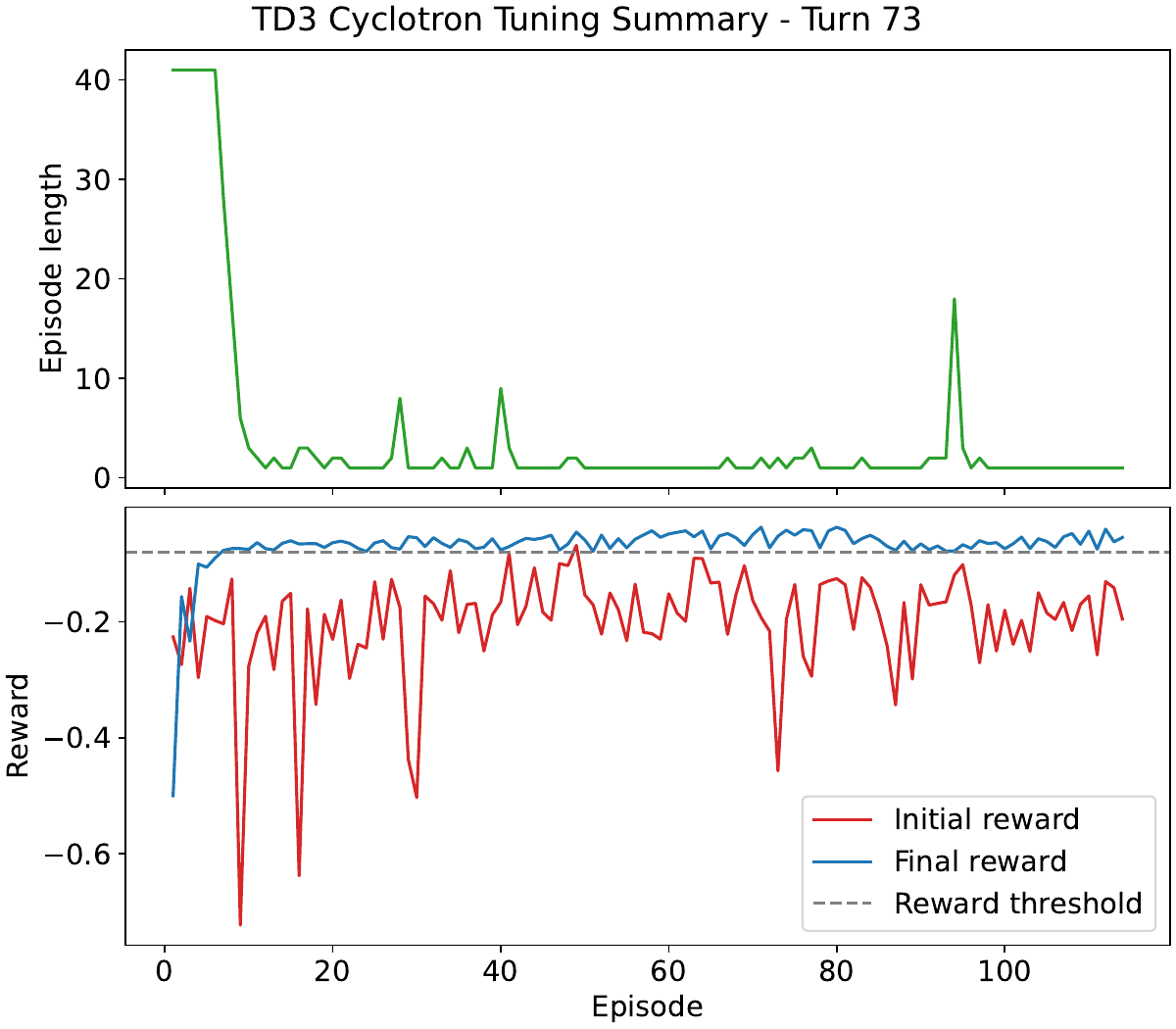}
\caption{Learning Curve for Turn 73 with pretraining: Top Plot, Episode Length during TD3 Training; bottom Plot, Reward evolution across episodes. Total training time is 460 timesteps, totaling 3.9 hours from start-to-end.}
\label{fig:reward-turn73}
\end{figure*}

Relative to Turn 72, the pretrained agent exhibited markedly faster convergence, requiring only about 2.3 hours (291 timesteps), while achieving equivalent final performance (average reward $\approx-0.06$). The learning curve in Fig.~\ref{fig:reward-turn73} shows a rapid stabilization of episode length to fewer than ten steps, with the reward reaching its steady value within the first few episodes. Phase alignment accuracy remained within $\pm1^{\circ}$ of the reference (Fig.~\ref{fig:phase-loss-stacks}, left column, row 2), losses stayed well below the 30 nA warning threshold (right column, row 2), and interlocks disappeared entirely after early exploration (Fig.~\ref{fig:coil-interlocks-stacks}, right column, row 2).

A direct comparison of the two training runs for Turn 73 confirmed the benefit of surrogate pretraining: the surrogate-initialized policy reached stable high rewards within a few episodes, whereas the agent trained from scratch exhibited larger fluctuations and slower convergence. Overall, surrogate-based pretraining provided a safer and more efficient initialization, enabling faster learning, smoother convergence, and stable operation throughout.

\textbf{Turn 74 (transfer from Turn 73):}  
This experiment tested policy transfer by initializing the TD3 agent with the actor network trained at Turn 73. Despite the warm start, the agent faced markedly different beam dynamics, in particular a smaller turn separation near extraction.

\begin{figure*}
\centering 
\includegraphics*[width=8cm]{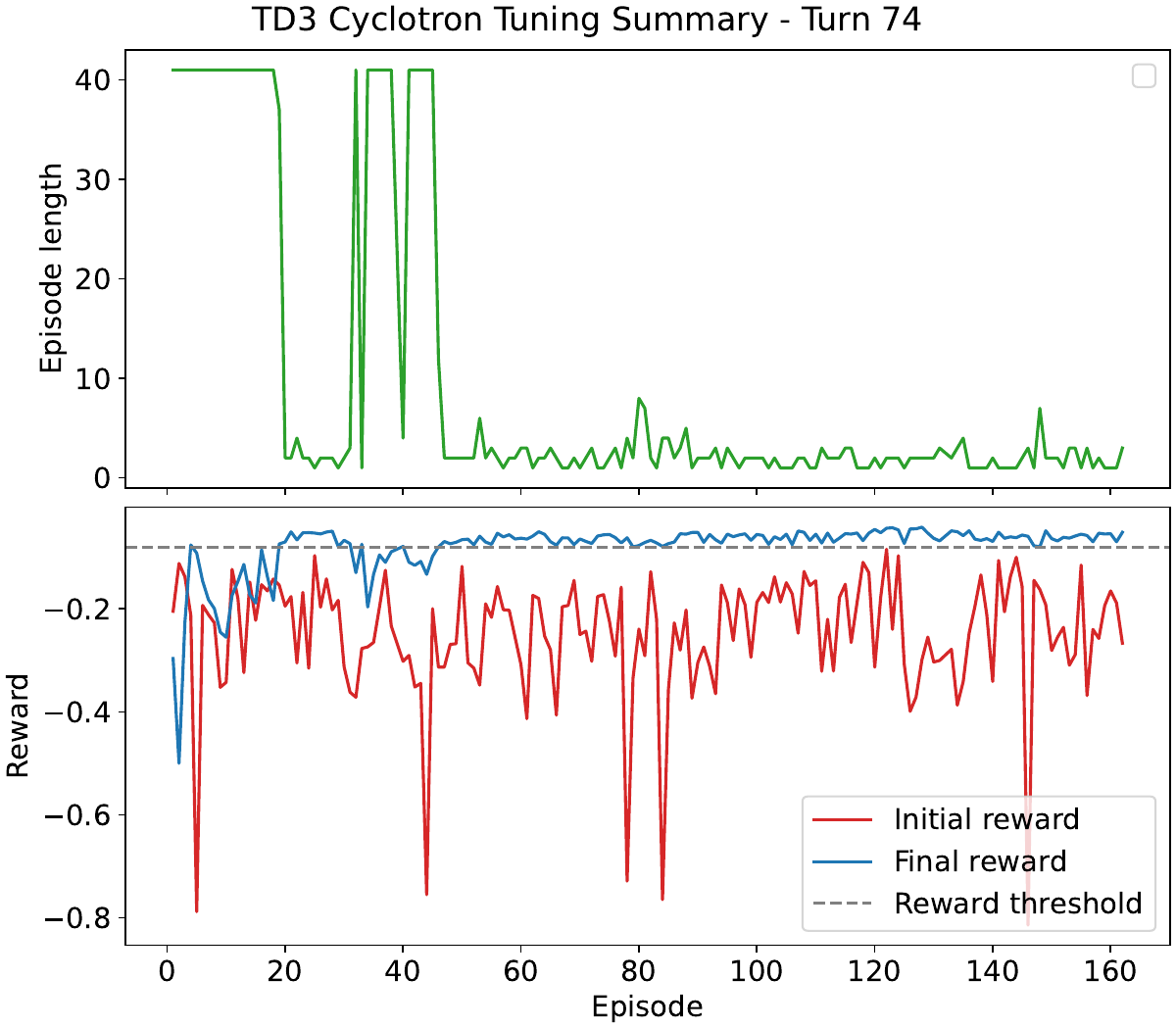}
\caption{Learning Curve for Turn 74: Top Plot, Episode Length during TD3 Training; bottom Plot, Reward evolution across episodes. Total training time is 1500 timesteps, totaling 7.8 hours from start-to-end.}
\label{fig:reward-turn74}
\end{figure*}

As a result, adaptation required roughly 50 episodes before the policy stabilized, leading to the longest overall convergence time among all investigated turns (5.8 hours, 1117 steps). The learning curve in Fig.~\ref{fig:reward-turn74} shows an initially slow improvement followed by steady convergence, with final rewards around $-0.06$. In addition, Turn 74 exhibited the highest number of interlocks among all runs, 51 in total, with three occurring after convergence (Fig.~\ref{fig:coil-interlocks-stacks}, right column, row 3).

Phase alignment accuracy remained within $\pm1.5^{\circ}$ of the reference (Fig.~\ref{fig:phase-loss-stacks}, left column, row 3), and beam losses, initially above 10 nA, were successfully reduced below 1 nA after convergence (right column, row 3), though overall loss suppression remained less effective than at Turn 72 and 73. Simultaneously, the mean trim-coil usage steadily decreased (Fig.~\ref{fig:coil-interlocks-stacks}, left column, row 3), indicating efficient control once adaptation was achieved.

Overall, while the agent ultimately achieved stable tuning and satisfactory reward performance, the extended adaptation time and elevated interlock rate highlight the limited generalization of policies across turns and the strong influence of underlying turn-dependent dynamics discussed in Sec.~\ref{sec:turn-current-dynamics}.

\begin{figure*}
\centering 
\includegraphics*[width=8cm]{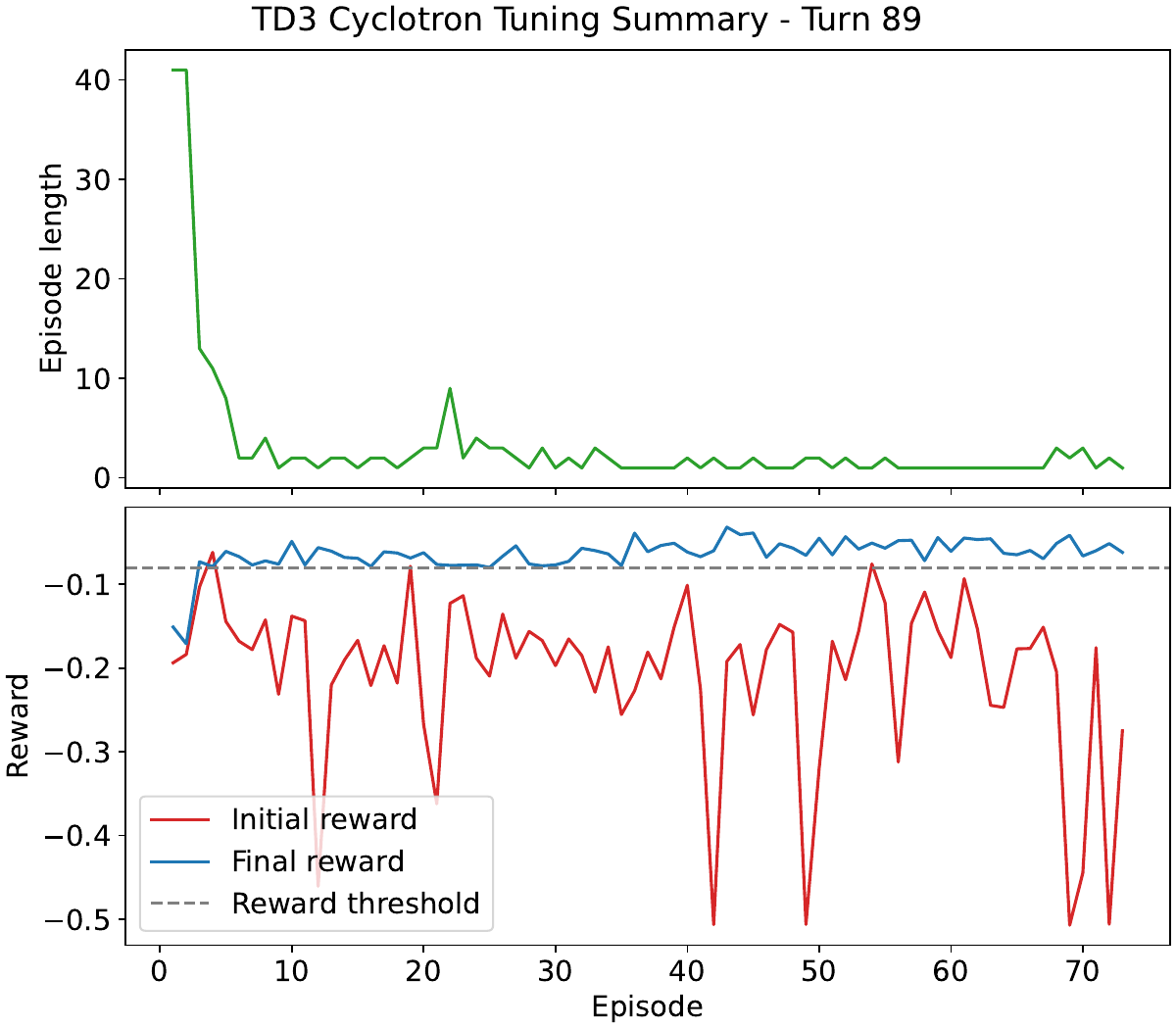}
\caption{Learning Curve for Turn 89: Top Plot, Episode Length during TD3 Training; bottom Plot, Reward evolution across episodes.}
\label{fig:reward-turn89}
\end{figure*}

\textbf{Turn 89 (comparison: surrogate pretraining vs. from scratch):}  
Turn 89 represented the most degraded configuration, with only two resonators active, resulting in smaller turn separation and high beam-loss sensitivity. As in Turn 73, surrogate pretraining was directly compared to training from scratch to evaluate the impact of historical-data initialization. Contrary to expectations, the pretrained agent exhibited unstable learning dynamics, larger reward fluctuations, and slower convergence toward the target threshold. In contrast, the agent trained from scratch achieved faster and more reliable convergence ($\sim$0.9 hours, 114 steps) without any interlocks (Fig.~\ref{fig:reward-turn89}).

Phase alignment accuracy remained within $\pm1.5^{\circ}$ of the reference (Fig.~\ref{fig:phase-loss-stacks}, left column, row 4), and beam losses were reduced by nearly two orders of magnitude, reaching sub-nA levels after convergence (right column, row 4). Coil activity stabilized accordingly, with no excessive actuation (Fig.~\ref{fig:coil-interlocks-stacks}, left column, row 4).

The poorer performance of the surrogate-pretrained model highlights the limitations of applying pretraining under substantially altered beam dynamics or machine setups. These findings emphasize that successful pretraining requires accurate regime matching between the surrogate domain and the target operational state.

\begin{figure*}
\centering 
\includegraphics*[width=8cm]{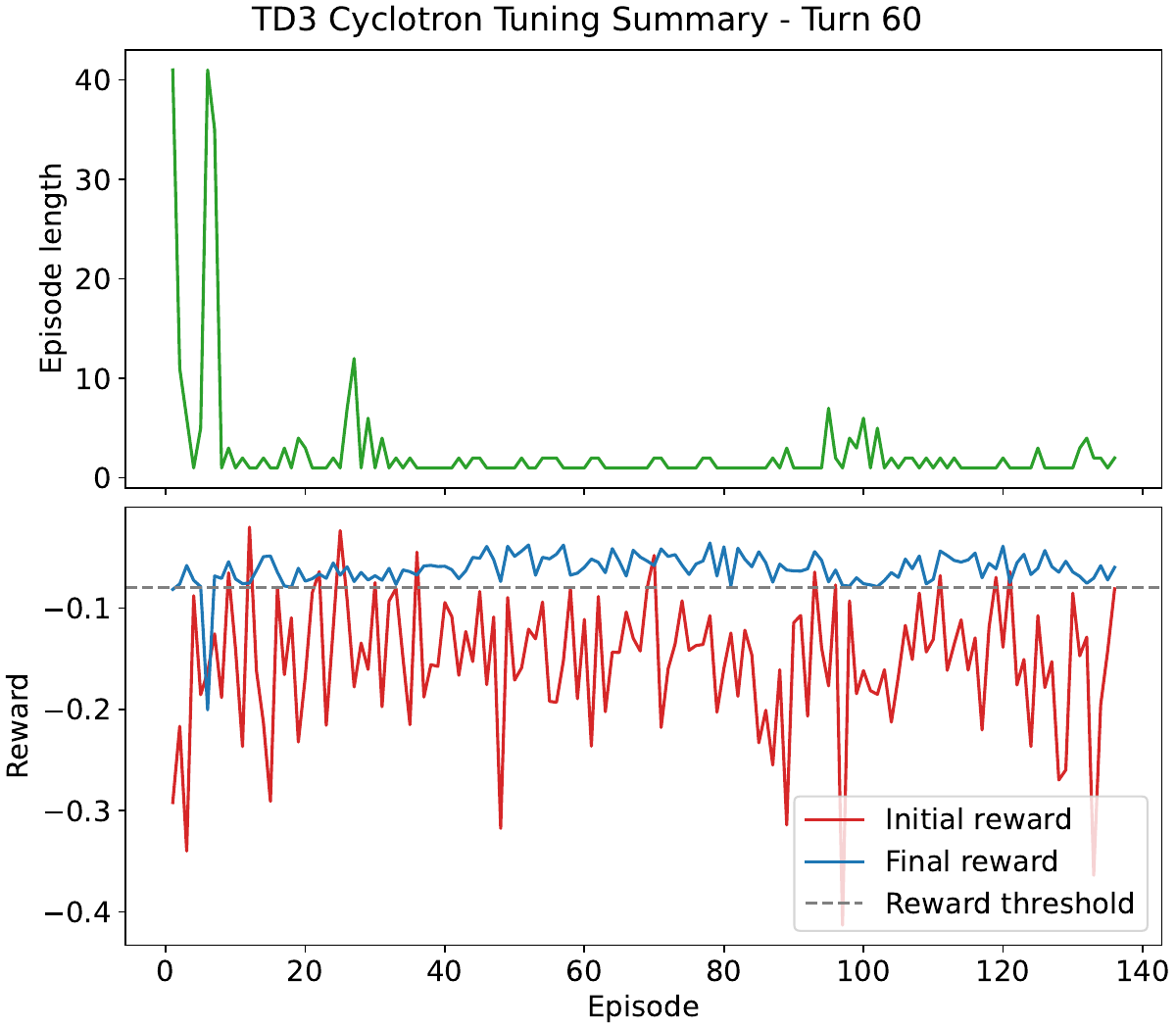}
\caption{Learning Curve for Turn 60: Top Plot, Episode Length during TD3 Training; bottom Plot, Reward evolution across episodes.}
\label{fig:reward-turn60}
\end{figure*}

\textbf{Turn 60 (nominal, all resonators):}  
The nominal configuration with all four resonators active provided the most stable operating regime and served as a benchmark for the RL framework. Without any pretraining, the agent converged in approximately 2.1 hours (217 steps) with only four interlocks, confirming the robustness and efficiency of online learning under nominal conditions.
The episode and reward evolution in Fig.~\ref{fig:reward-turn60} demonstrate rapid convergence within fewer than 20 episodes, with smooth and stable final rewards. Phase alignment accuracy remained within $\pm1.5^{\circ}$ of the reference (Fig.~\ref{fig:phase-loss-stacks}, left column, row 5), while beam losses steadily decreased and remained well below operational thresholds after convergence (right column, row 5). Trim-coil usage was consistently low (Fig.~\ref{fig:coil-interlocks-stacks}, left column, row~5), indicating that the agent benefited from the larger turn separation to maintain high beam quality with minimal corrective intervention. These results establish Turn 60 as a strong baseline for reliable RL-based tuning and confirm that the learning framework performs most effectively under well-separated beam dynamics.

\textbf{Summary and Outlook.}
Across all investigated turn configurations, the TD3-based RL framework consistently achieved stable and reproducible tuning of the Injector~2 cyclotron within a few hours of online training (typically under 1000 timesteps). The experiments demonstrated that, once equipped with an appropriate reward formulation and interlock-aware safety layer, the RL agent was capable of reliably aligning beam phases, suppressing losses, and minimizing corrective actions across varying operational regimes.
Pretraining using surrogate data proved beneficial under comparable machine conditions (e.g., Turn~73), but introduced bias under strongly altered dynamics (Turn~89), highlighting the importance of accurately matching simulation and real-machine behavior. Furthermore, policy transfer between turns (Turn~73~$\rightarrow$~74) was found to be limited, confirming that turn-dependent beam dynamics require localized adaptation.
Overall, these results establish RL as a viable and safe control paradigm for real-time cyclotron tuning.

\subsection{Overnight Evaluation and Reliability Testing}
To assess the reliability and robustness of the trained RL policy under realistic accelerator operation, the TD3 agent was deployed in evaluation mode on the Injector~2 cyclotron during a series of overnight runs conducted after daytime training sessions. In this mode, the agent acted in a purely deterministic manner, executing its learned policy few hours earlier without further gradient updates or exploration noise, thus directly testing its capacity for autonomous stabilization and long-term control.

The evaluation was performed across three representative configurations: Turn~74 (three resonators active), Turn~89 (two resonators active), and Turn~60 (four resonators active). These settings span increasing levels of operational stability and beam-loading complexity, providing a stringent test of the policy’s generalization capability. During each overnight session, the agent retained full control of the trim coils, main correction coil (AIHS), and resonator voltage (CI3V). To emulate realistic machine drifts, small random perturbations (“kicks”) were occasionally introduced in the actuator setpoints, allowing assessment of the agent’s ability to autonomously recover optimal beam conditions.

\subsubsection{Agent Response to Drifts and Perturbations}
\begin{figure*}[t]
    \centering
    \begin{minipage}[t]{0.48\textwidth}
        \centering
        \includegraphics[width=0.8\linewidth]{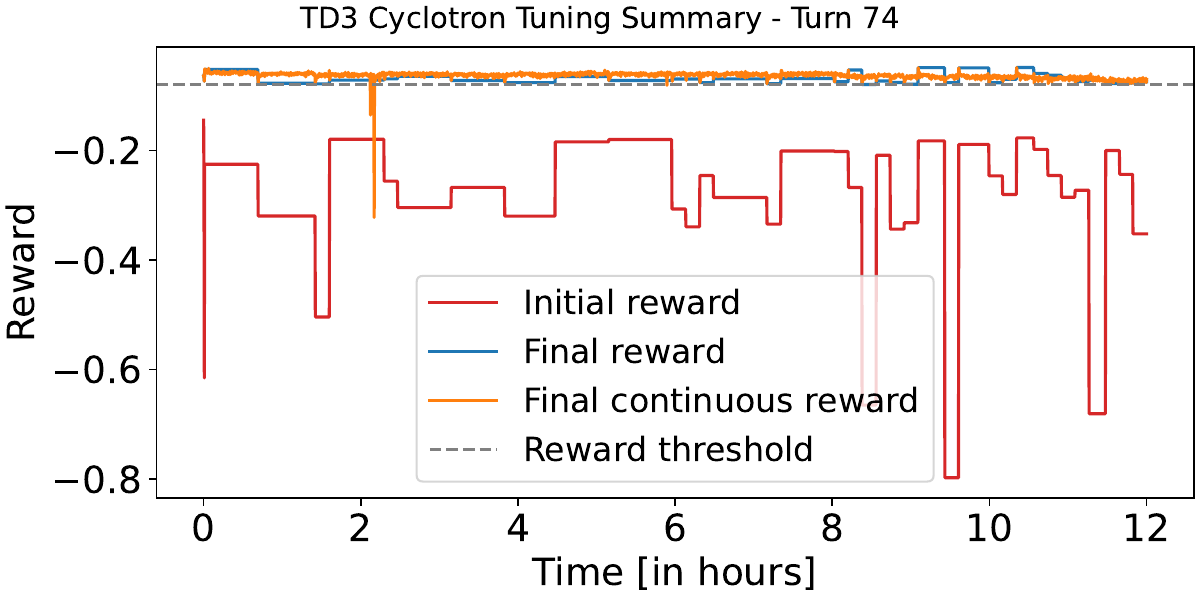}\\[0.8em]
        \includegraphics[width=0.8\linewidth]{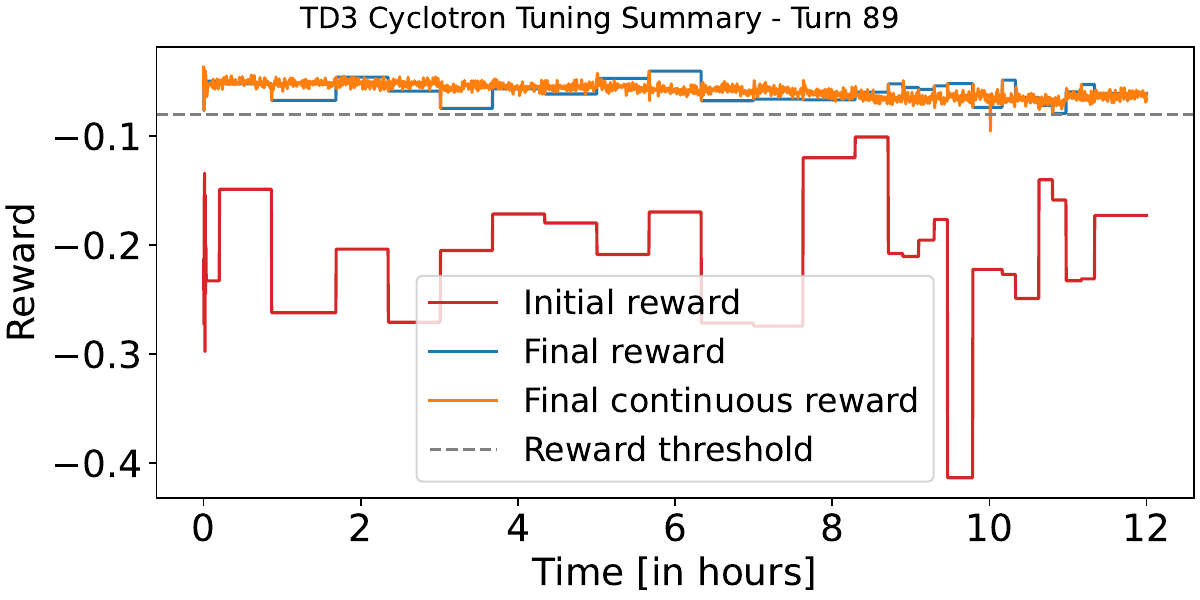}\\[0.8em]
        \includegraphics[width=0.8\linewidth]{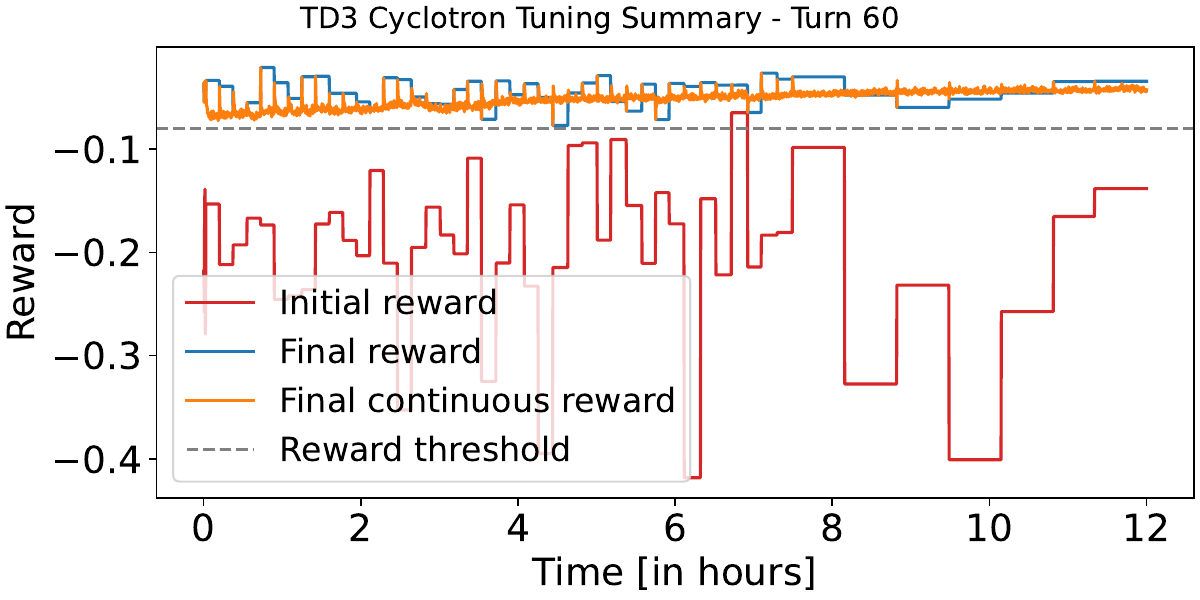}
    \end{minipage}\hfill
    \begin{minipage}[t]{0.48\textwidth}
        \centering
        \includegraphics[width=0.8\linewidth]{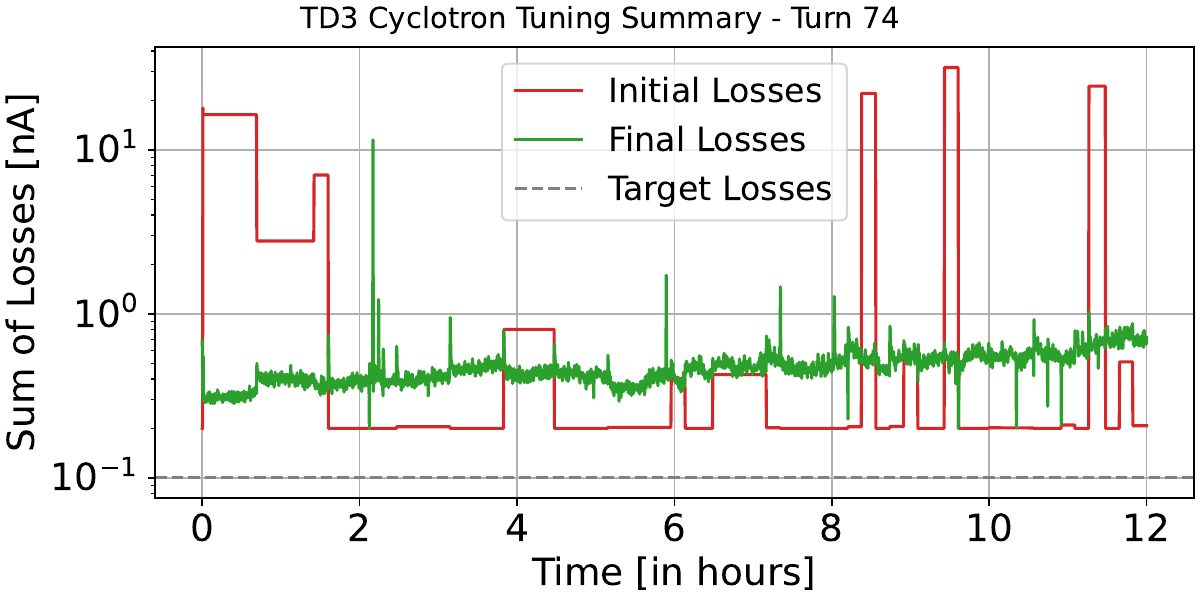}\\[0.8em]
        \includegraphics[width=0.8\linewidth]{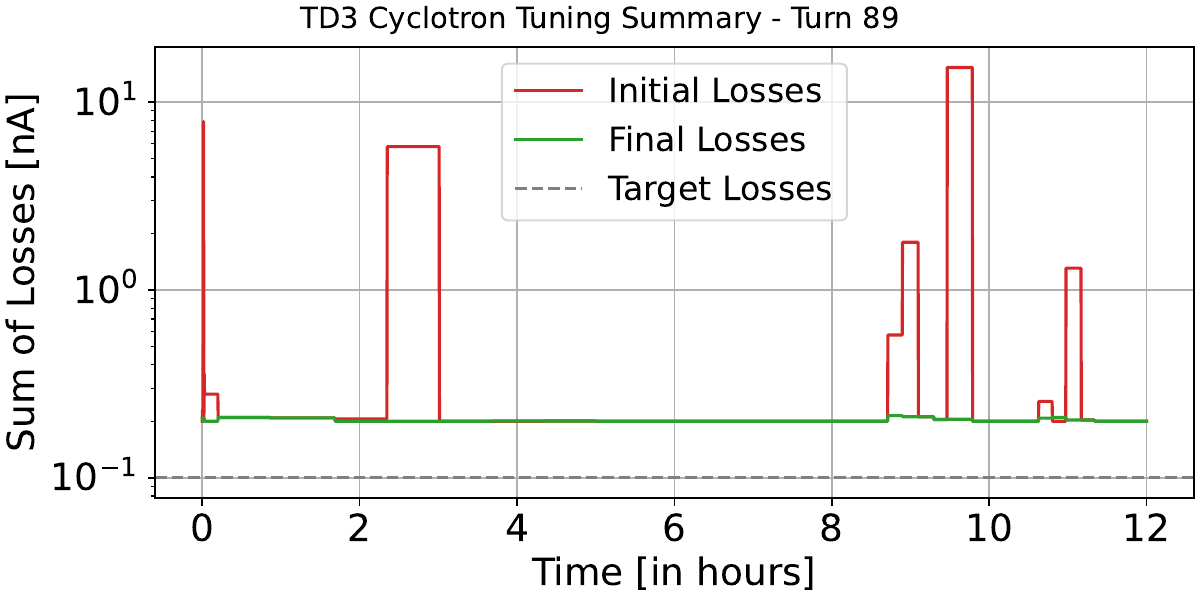}\\[0.8em]
        \includegraphics[width=0.8\linewidth]{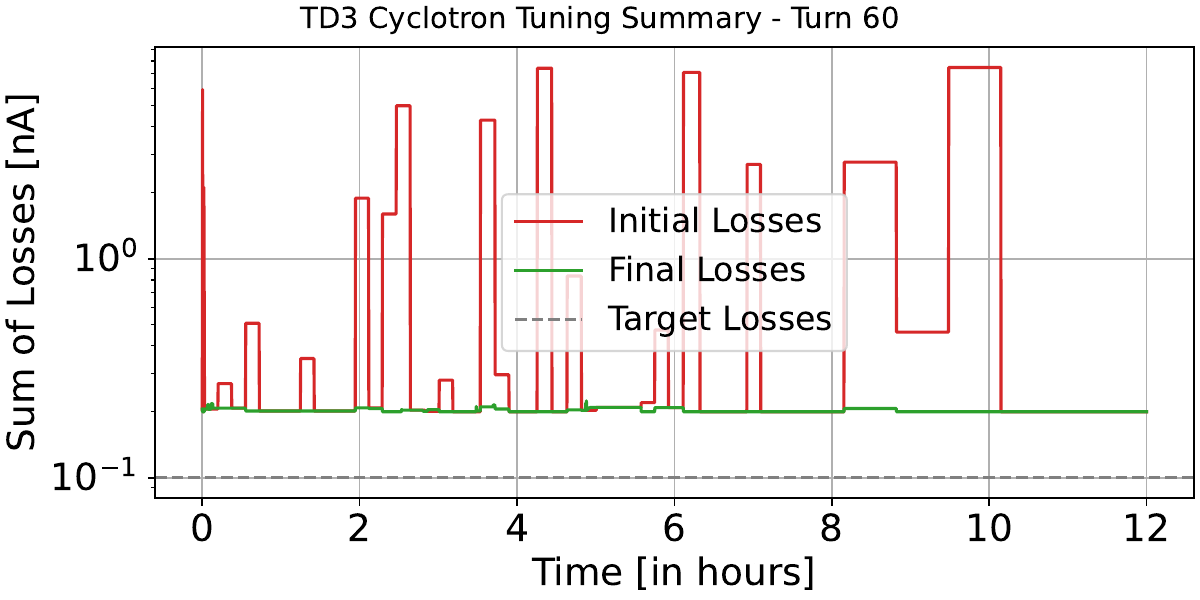}
    \end{minipage}

    \caption{Overnight evaluation of the trained TD3 policy across three representative configurations (top to bottom: Turns~74, 89, and~60). 
    Left column: evolution of initial and final rewards during continuous, autonomous operation. 
    Right column: corresponding beam loss evolution over the same intervals. 
    In all cases, the agent successfully maintained high-reward, low-loss operation, autonomously compensating for perturbations without triggering interlocks.}
    \label{fig:overnight-reward-loss}
\end{figure*}

Figure~\ref{fig:overnight-reward-loss} summarizes the evolution of the initial and final rewards throughout the evaluation for Turns~74, 89, and~60. Across all configurations, the agent consistently maintained stable operation and promptly corrected perturbations, restoring the system to high-reward conditions without triggering any interlocks.

For Turn~74, the first configuration tested, the reward threshold was conservatively kept at its training value of $-0.08$, while for Turns~60 and~89, the threshold was tightened to $-0.06$ to further penalize beam losses and incentivize the agent to improve its tuning. In every case, the final reward remained above the threshold, demonstrating the policy’s ability to maintain beam quality despite external disturbances and drifts. The difference between initial and final rewards illustrates the closed-loop corrective response, confirming that the trained RL policy can stabilize the cyclotron in real time under natural drifts and hardware fluctuations.

\subsubsection{Interlock Behavior and Recovery Performance}
The evolution of beam losses during these tests is shown in Fig.~\ref{fig:overnight-reward-loss}.
For Turn~74, a gradual increase in beam losses was observed overnight, although no interlocks were triggered and the system remained under stable control. A brief retraining session (approximately 30~minutes) performed the following day successfully restored losses to near-zero levels, demonstrating that short, targeted updates suffice to recover optimal performance.

In contrast, for Turns~60 and~89, where the higher reward threshold of $-0.06$ was applied, the agent maintained both low beam losses and stable coil usage throughout the evaluation. These results confirm that the trained RL policy not only preserves safe operation over extended periods but also autonomously compensates for small disturbances without human intervention.

\subsubsection{Discussion}
The overnight evaluation results confirm that the RL-based control framework generalizes reliably beyond its training conditions.
The policy demonstrated strong resilience to environmental drifts, maintained beam stability across a range of operating regimes, and operated fully autonomously for several hours without triggering interlocks.
Furthermore, the ability to restore performance via brief retraining underscores the value of intermittent fine-tuning as a sustainable long-term operation strategy. \\
It is important to emphasize that the present work was not intended to replace expert operators during the initial setup of a new machine configuration. In the experiments reported here, operator-led turn changes were generally completed faster than online RL training. Rather, the objective was to learn a state-dependent control policy that could subsequently operate autonomously under realistic machine conditions. In this context, the most significant result is the successful deployment of trained policies during extended evaluation periods, where the agent compensated operational drifts, recovered from perturbations, and maintained stable beam conditions without further learning.

These findings highlight the readiness of RL-based tuning for deployment in continuous operation modes of high-power proton cyclotrons, where reliability, adaptability, and safety are critical.

\subsection{Generalization to Higher Current}

The final phase of the campaign assessed the ability of the RL model, trained exclusively under low-current conditions at 20 µA, to generalize when evaluated at higher beam currents. Experiments were conducted with all four resonators active, corresponding to a 60-turn configuration representing nominal operation conditions. The beam current was systematically increased following a Fibonacci-like ramp-up sequence (100, 200, 300, 500, and 800 µA), within a continuous evaluation interval of approximately 1.5 hours.

At every current level, the system was first perturbed by random variations of actionable parameters within the predefined exploration space. The RL agent was then tasked with restoring optimal beam conditions by maximizing the reward function, which implicitly minimized beam losses. As in previous stages of the campaign, success was defined by surpassing the reward threshold of -0.08, after which the agent continued monitoring and fine-tuning. Across all high-current tests, the RL agent restored the beam in at most five steps and no interlocks were triggered, confirming its robustness and safety. A summary of results is provided in Table~\ref{tab:high_current}.

\begin{table*}[htbp]
    \centering
    \caption{Performance of the RL model at increasing beam currents.}
    \label{tab:high_current}
    \begin{tabular}{cccccc}
    \hline
    Current (µA) & Initial Reward & Final Reward & Initial Losses (nA) & Final Losses (nA) & Adjustment Steps \\
    \hline
    50  & -0.14 & -0.041 & 0.2   & 0.2 & 1 \\
    100 & -0.25 & -0.046 & 2.0   & 0.2 & 3 \\
    200 & -0.12 & -0.049 & 0.2   & 0.2 & 1 \\
    300 & -0.77 & -0.060 & 95.0  & 0.3 & 3 \\
    500 & -0.33 & -0.062 & 7.0   & 0.2 & 3 \\
    800 & -0.54 & -0.069 & 23.3  & 0.3 & 5 \\
    \hline
    \end{tabular}
\end{table*}

As the current exceeded 300 µA, modest but expected complications were noted. Beam size and position shifts are inherent to the physics of increasing beam current. These effects made the system more sensitive to perturbations and rendered the random reset procedure less straightforward. In some instances, unsafe parameter combinations from the reset logic triggered interlocks before the RL agent could act. Although not a major issue, this required attention: the action space was narrowed and the maximum permissible step size reduced, enabling stable recovery while preserving adaptability.

In summary, the results demonstrate that the RL model generalizes reliably to higher-current conditions up to the maximum tested level of 800 µA. The increased beam sensitivity highlights the importance of adaptive action space tuning and parameter scaling when extending RL-based control to higher-intensity regimes.

\section{Conclusion}
The present study demonstrates, for the first time, that reinforcement learning (RL) can be reliably deployed for autonomous closed-loop tuning of a high-power cyclotron. Through a series of controlled experiments on Injector~2 at PSI, the TD3 agent successfully optimized beam phase and losses across five turn configurations, encompassing both nominal and degraded operating regimes.
These results show that ML-based control can operate within strict safety boundaries while achieving convergence within a few hours, even under varying machine conditions.

From a methodological standpoint, the experiments addressed several fundamental questions regarding algorithm design, observability, and transferability.
The TD3 framework proved particularly suitable for this task, striking a balance between sample efficiency and operational safety in a multi-dimensional continuous control space.
The use of phase and loss signals as state variables provided an effective, physically meaningful representation of the beam-machine interaction.
Action-space analysis confirmed that AIHS (main field) and CI3V (cavity voltage) dominated the beam phase response, while the trim coils served as fine-correction channels, enabling efficient control with minimal intervention.

Pretraining emerged as a valuable, though context-dependent acceleration strategy.
When the surrogate data reflected conditions close to the real setup (as at Turn~73), convergence was significantly faster and safer than from random initialization.
Conversely, in degraded configurations (Turn~89), pretraining introduced a bias that impeded adaptation, underscoring the need for careful alignment between surrogate models and the real operating regime.
This interplay between model fidelity and learning efficiency highlights the importance of combining physics-based surrogate models with adaptive, online retraining for robust deployment.

Across all configurations, convergence trends mirrored those predicted by single-particle simulations, validating the use of physics-guided RL environments for pre-experiment benchmarking.
Furthermore, policy robustness was confirmed during overnight evaluation runs, where the trained agents maintained beam stability and corrected imposed perturbations autonomously, without triggering interlocks.
This establishes not only the feasibility but also the repeatability of ML-based tuning under real operating conditions.

Importantly, the campaign revealed that policies do not generalize across turn numbers without adaptation. This is fully consistent with the turn-dependent beam dynamics discussed in Sec.~\ref{sec:turn-current-dynamics}.
However, warm-starting from previous policies and targeted retraining substantially reduced adaptation time, pointing toward the viability of hierarchical or multi-turn RL frameworks for future deployment.

A further step toward operational generalization was demonstrated through high-current tests, where the RL policy trained exclusively at 20 µA maintained stable and low-loss operation up to 800 µA ($\approx$ 58 kW average beam power) without additional training.
The agent consistently restored optimal beam conditions in fewer than five steps at all current levels, without triggering interlocks.
At higher currents, increased beam sensitivity introduced expected operational challenges, mainly related to reduced stability margins and the increased impact of random perturbations. These effects were successfully mitigated through tighter action constraints and adapted step-size limits, while preserving the effectiveness of the learned policy.
These results demonstrate successful transfer of a low-current-trained policy across the full current range investigated in this study. Although operational constraints were adjusted to accommodate the increased beam sensitivity at higher currents, the same trained policy remained effective throughout the current-scaling experiments without retraining.

It is important to note that the present work does not demonstrate recovery from major hardware failures on the timescale required for ADS operation. Rather, it demonstrates autonomous compensation of operational drifts and perturbations over timescales of seconds to hours, representing a critical intermediate step toward future fault-tolerant accelerator control systems.

Looking forward, these findings provide a solid foundation for scaling the approach to the full HIPA complex and, ultimately, to ADS.
The path toward HIPA integration will follow a staged, safety-certified roadmap: (i) simulation-first validation of each subsystem, (ii) modular agent design combining operational and safety controllers, and (iii) incremental deployment under operator supervision.
The broader goal is to establish a transparent, certifiable ML control architecture capable of sustaining long-term, high-current operation with minimal human intervention.
Such a system would not only enhance operational reliability for HIPA but also address the stringent autonomy and fault-tolerance requirements of ADS-class accelerators.

In summary, this work provides compelling evidence that reinforcement learning can transition from simulation to real-machine control in a safe, interpretable, and operationally meaningful way.
By coupling physical modeling with adaptive optimization, the approach paves the way for intelligent, self-correcting accelerator control systems, an essential step toward the next generation of high-power proton drivers.

\begin{acknowledgments}
The authors acknowledge the support of the PSI management and the Transmutex management for enabling and supporting this experimental campaign. We thank Christian Baumgarten, Mariusz Sapinski, Luciano Calabretta, Rudolf D\"{o}lling, and Jilei Sun for valuable discussions. Special thanks are extended to the PSI control room crew for their essential support during the Injector~2 experimental campaign. The main author also wishes to thank Verena Kain and Michael Schenk for insightful discussions and feedback during the CERN School of Computing, which helped shape several aspects of this work.
\end{acknowledgments}

\clearpage
\appendix

\section{Correlation Heatmap}\label{app:heatmap}
\begin{figure}[htbp]
    \centering
    \includegraphics[width=0.95\linewidth]{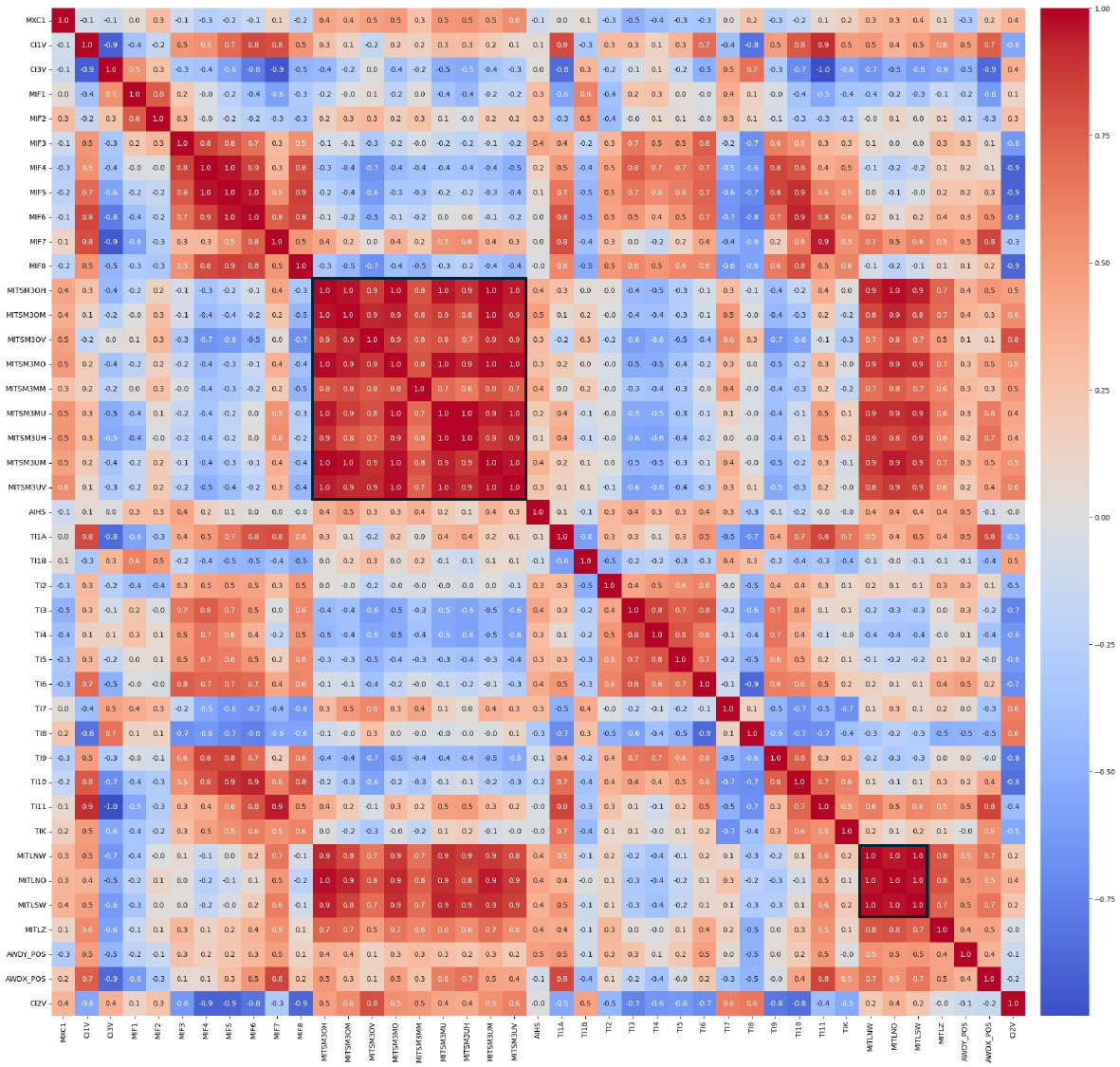}
    \caption{Main features correlation heatmap (temperature sensors correlations highlighted).}
    \label{fig:heatmap}
\end{figure}

For convenience, Table~\ref{tab:abbreviations} summarizes the principal Injector~2 instrumentation and abbreviations referenced throughout the manuscript and used in the correlation matrix shown in Fig.~\ref{fig:heatmap}.
\begin{table*}[ht]
\centering
\caption{Principal Injector~2 instrumentation and abbreviations used throughout this work.}
\label{tab:abbreviations}
\renewcommand{\arraystretch}{1.2}
\begin{tabular}{lll}
\hline
\textbf{Abbreviation} & \textbf{Description} & \textbf{Role in this work} \\
\hline
AIHS & Main correction magnet current & Global beam phase adjustment \\
TI1--TI12 & Trim-coil currents & Local magnetic field correction \\
CI3V & Resonator 3 peak voltage & RF voltage optimization \\
MIF1--MIF8 & Beam phase probes & Beam phase measurements (reward and state) \\
MXC1 & Extracted beam current monitor & Beam transmission and stability \\
MII7 & Ionization chamber & Beam-loss monitor near the extraction \\
MXI1 & Ionization chamber & Beam-loss monitor at the subsequent beamline \\
MITL & Mean air temperature & Environmental state variable \\
MITSM & Mean magnet temperature & Magnet thermal state \\
RIE1 & Beam Profile at extraction & Radial beam distribution \\
\hline
\end{tabular}
\end{table*}

\section{Historical Data Preparation}\label{app:data}
Historical operational data were extracted from the PSI archival system using dedicated Python tools developed at PSI. The archived channels originate primarily from Injector~2, complemented by downstream beam-loss monitors and extracted beam-current measurements.

Because different process variables are archived at different sampling frequencies, all channels were resampled to a uniform interval of 200~ms using the most recent available value. This produced a synchronized dataset suitable for machine-learning applications.

To ensure consistency between historical operation and the experimental campaign, only periods corresponding to representative resonator configurations were retained:

\begin{itemize}
\item May 2023 -- December 2023:
two double-gap resonators with a low-voltage flattop cavity.

\item May 2024 -- July 2024:
two double-gap resonators.

\item August 2024 -- December 2024:
two double-gap resonators and one single-gap resonator.
\end{itemize}

A held-out dataset was extracted from the central portion of each filtered dataset. Approximately one million samples from each operating period were reserved for validation and testing. \\
To compensate for the unequal duration of the operating periods, the larger datasets were down-sampled prior to training. The final training dataset contained approximately 27 million samples, corresponding to more than 63 days of Injector~2 operation.

\section{Surrogate Model Architecture and Training} \label{app:surrogate}
Several surrogate-model candidates were evaluated, and the final selected model was a Mixture Density Network (MDN) implemented in PyTorch. The network consisted of an input layer of size 21 followed by three fully connected hidden layers with 256, 128, and 128 neurons, respectively. Batch normalization was applied after the first two hidden layers. Feature weighting was introduced at the network input to emphasize the most influential control variables. Most input features were assigned a weight of 0.1, whereas the main-coil current (AIHS) and the resonator~3 voltage (CI3V) were assigned weights of 1.0 and 0.5, respectively, reflecting their dominant influence on the beam phase response. The loss function also employed nonuniform target weights for the MIF signals, $[1,1,1,1,2,3,4,5]$, giving progressively greater importance to downstream phase probes closer to extraction.

The MDN was trained by minimizing a weighted negative log-likelihood loss associated with the Gaussian-mixture output. Training was performed for 23 epochs using a batch size of 1024. The final model checkpoint was selected according to the minimum validation mean-squared error. On the held-out test dataset, the selected model achieved an average mean-squared error of $2.4\times10^{-5}$ and an average coefficient of determination of $R^2=0.9898$, demonstrating that the surrogate accurately reproduced the measured MIF phase response over the historical operating range.

\section{Jacobian analysis and sensitivity evolution}\label{app:jacobian}
To quantify how control actuators affect the phase probes, turn-dependent Jacobian matrices were measured using finite-difference perturbations \cite{parfenova2016beamphase}. Each control parameter (AIHS, CI3V, and the twelve trim coils) was perturbed individually, with stabilization periods inserted between perturbations to ensure reproducible phase readings. The resulting Jacobians capture the partial derivatives of the measured phases (MIF1–MIF8) with respect to each actuator.

\begin{figure}
\centering 
\includegraphics*[width=10cm]{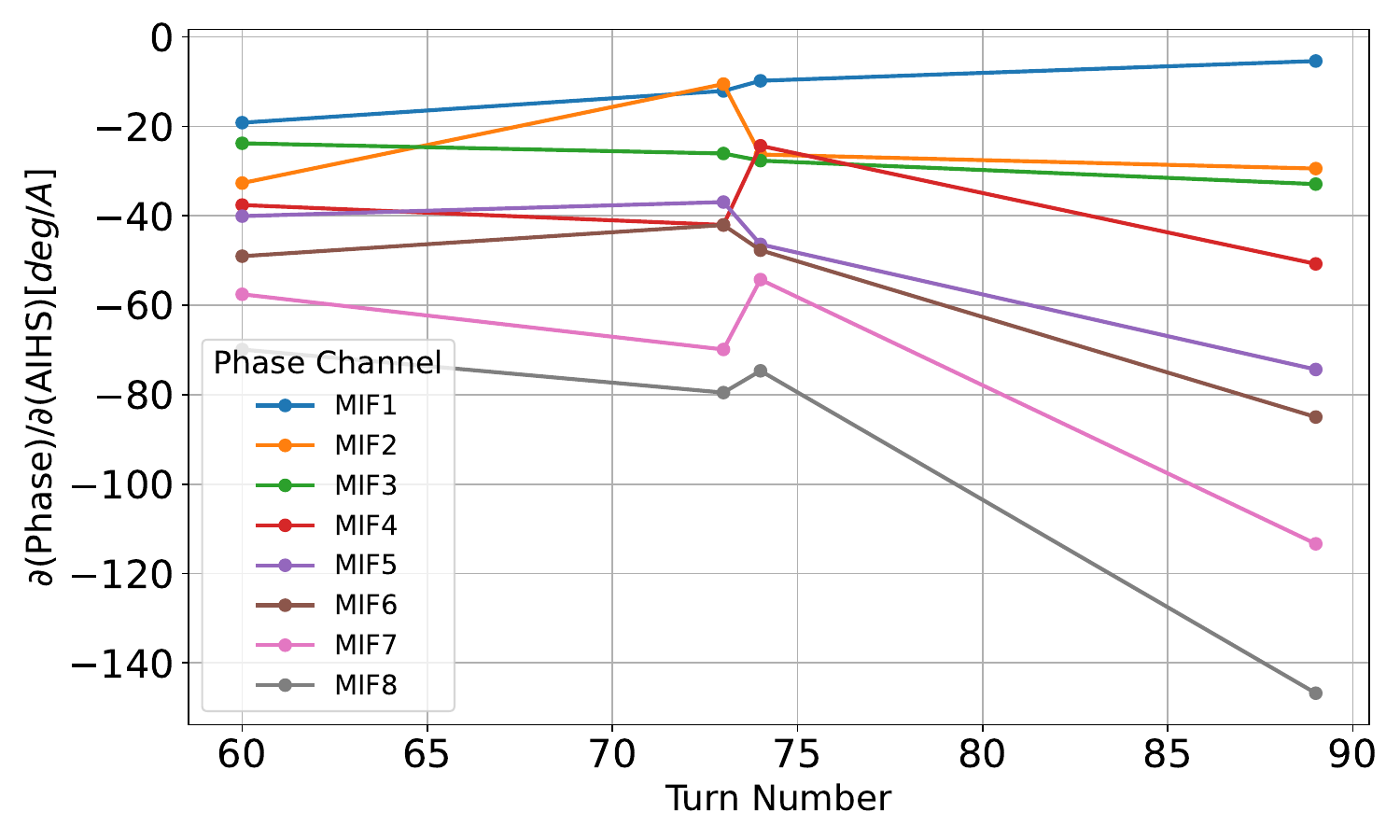}
\caption{Sensitivity of phases to AIHS change with respect to the turn number, expressed in degrees per ampere (deg/A), for a perturbation amplitude of 0.02~A.}
\label{fig:AIHS_sensitivity}
\end{figure}

The measured Jacobians exhibited strong dependence on the turn number. In particular, the sensitivity of the outer probes (MIF4–MIF8) to changes in AIHS increased markedly at higher turn numbers. This behavior reflects the underlying beam dynamics: reducing the voltage of resonators~2 and~4 forces the beam to take additional turns at large radii, where changes in the main magnetic field have a magnified effect. Conversely, early probes (MIF1–MIF3) showed only weak dependence, consistent with their location near injection. Similar trends were observed for specific trim coils, such as TI11 and TI10, whose influence on the phase distribution varied significantly with turn number. Representative sensitivities are shown in Fig.~\ref{fig:AIHS_sensitivity}.

Beyond turn dependence, the Jacobians also displayed strong sensitivity to the working point. Even for fixed perturbation amplitudes, the measured responses varied significantly depending on the resonator and coil settings chosen to achieve a given turn number. This working-point dependence reflects the nonlinear character of the machine and emphasizes the need for adaptive, data-driven control strategies.



\nocite{*}

\clearpage
\bibliography{main}

\end{document}